\pdfoutput=1
\documentclass[pdflatex,sn-mathphys-num,iicol]{sn-jnl}

\usepackage{graphicx}
\usepackage{amsmath,amssymb,amsfonts}
\usepackage{xcolor}
\usepackage{booktabs}

\usepackage{subcaption}

\usepackage[ruled,vlined]{algorithm2e}

\usepackage{bm}

\usepackage{tikz}
\usetikzlibrary{calc,shapes.geometric}

\usepackage{hyperref}
\usetikzlibrary{svg.path}
\definecolor{authorcidcol}{HTML}{A6CE39}
\tikzset{authorcidlogo/.pic={
  \fill[authorcidcol] svg{M256,128c0,70.7-57.3,128-128,128C57.3,256,0,198.7,0,128C0,57.3,57.3,0,128,0C198.7,0,256,57.3,256,128z};
  \fill[white] svg{M86.3,186.2H70.9V79.1h15.4v48.4V186.2z}
               svg{M108.9,79.1h41.6c39.6,0,57,28.3,57,53.6c0,27.5-21.5,53.6-56.8,53.6h-41.8V79.1z M124.3,172.4h24.5c34.9,0,42.9-26.5,42.9-39.7c0-21.5-13.7-39.7-43.7-39.7h-23.7V172.4z}
               svg{M88.7,56.8c0,5.5-4.5,10.1-10.1,10.1c-5.6,0-10.1-4.6-10.1-10.1c0-5.6,4.5-10.1,10.1-10.1C84.2,46.7,88.7,51.3,88.7,56.8z};
}}
\newlength{\authorcidXh}
\DeclareRobustCommand\authorcidlink[1]{%
  \setlength{\authorcidXh}{\fontcharht\font`X}%
  \href{https://orcid.org/#1}{\mbox{%
    \begin{tikzpicture}[yscale=-0.00390625*\authorcidXh,xscale=0.00390625*\authorcidXh,transform shape]
    \pic{authorcidlogo};
    \end{tikzpicture}}}}
\makeatletter
\newcommand{\authorORCID}[1]{\g@addto@macro\artauthors{\hskip2pt\authorcidlink{#1}}}
\makeatother

\AtBeginDocument{%
  \captionsetup{font=small}      %
  \captionsetup[sub]{font=footnotesize} %
}

\newcommand{\sota}[1]{\textbf{\textcolor{red}{#1}}}

\newcommand{\greenstar}{%
  \tikz[baseline=-0.6ex]%
    \node[
      star,
      star points=5,
      star point ratio=2.3,   %
      draw=black,
      fill=green!60!black,    %
      minimum size=5.5pt,
      inner sep=0pt
    ]{};
}

\newcommand{\redstar}{%
  \tikz[baseline=-0.6ex]%
    \node[
      star,
      star points=5,
      star point ratio=2.3,   %
      draw=black,
      fill=red,
      minimum size=5.5pt,
      inner sep=0pt
    ]{};
}

\tikzset{
  modlabel/.style={
    fill=white, fill opacity=.70, text opacity=1,
    rounded corners=2pt, inner sep=2pt,
    font=\scriptsize
  }
}

\begin{document}

\title[PBSR: Pseudo-Burst Super-Resolution via Zero-Shot Novel View Synthesis]{PBSR: Pseudo-Burst Super-Resolution via Zero-Shot Novel View Synthesis}

\author*[1,2]{\fnm{Yuki} \sur{Kondo}}\authorORCID{0000-0002-5263-8722}\email{yuki\_kondo\_ab@mail.toyota.co.jp}

\author[1]{\fnm{Norimichi} \sur{Ukita}}\authorORCID{0000-0002-0240-1065}\email{ukita@toyota-ti.ac.jp}

\affil[1]{\orgname{Toyota Technological Institute},
  \orgaddress{\street{2-12-1 Hisakata, Tenpaku-ku}, \city{Nagoya}, \state{Aichi}, \postcode{468-8511}, \country{Japan}}}

\affil[2]{\orgname{Frontier Research Center, Toyota Motor Corporation},
  \orgaddress{\street{543, Kirigahora, Nishihirose-cho}, \city{Toyota}, \state{Aichi}, \postcode{470-0309}, \country{Japan}}}

\abstract{Burst super-resolution (BSR) achieves higher reconstruction fidelity than single-image super-resolution (SISR) by aggregating sub-pixel information from multiple frames. However, it requires a burst of frames captured under near-static conditions, and such frames are unavailable for single photographs or archival images. We propose Pseudo-Burst Super-Resolution (PBSR), which brings BSR-style reconstruction to the single-image input setting by synthesizing pseudo-burst frames via zero-shot novel view synthesis (NVS). PBSR first generates pseudo-burst frames with realistic parallax using a viewpoint sampling strategy calibrated to the pixel-shift distribution of real burst datasets. The frames are then processed by a BSR backbone equipped with a geometry-aware alignment module that reuses the NVS camera parameters and estimated depth. While the NVS stage is zero-shot, the backbone is fine-tuned with an artifact-aware adaptation so that it can exploit the synthesized frames despite their generative artifacts. Experiments on the BurstSR dataset show that PBSR consistently improves in PSNR, SSIM, and LPIPS over the single-frame baseline with the same reconstruction backbone.
}

\keywords{Burst super-resolution, Single-image super-resolution, Novel view synthesis, Geometric alignment, Pseudo-burst generation, Domain adaptation}

\maketitle

{\renewcommand{\thefootnote}{}\footnotetext{Accepted for publication in \textit{New Generation Computing}. DOI: \href{https://doi.org/10.1007/s00354-026-00336-6}{10.1007/s00354-026-00336-6}}}

\section{Introduction}
\label{sec:intro}

\begin{figure}[htb]
  \centering
  \scriptsize
  {\centering
  \begin{subfigure}[t]{0.48\linewidth}
    \centering
    \includegraphics[width=\linewidth, trim=0 30 0 110, clip]{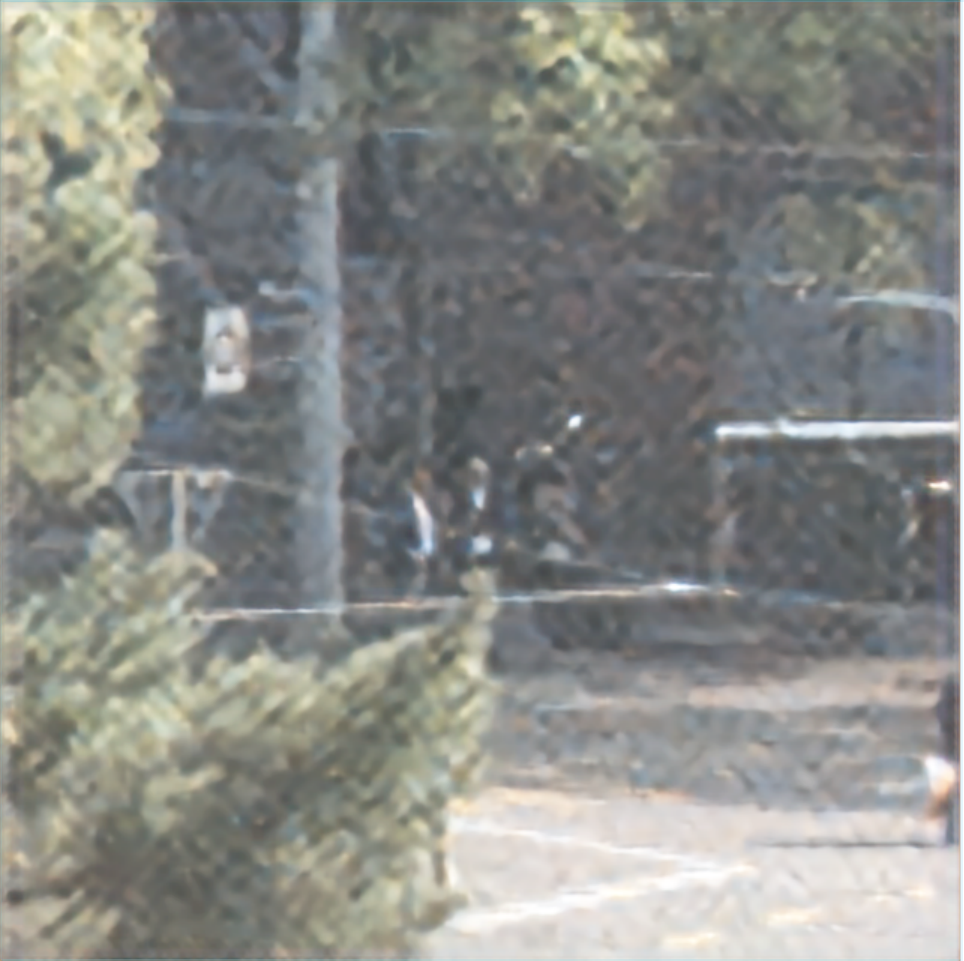}
    \caption{SISR: PSNR=51.09}
    \label{fig:ex_sisr}
  \end{subfigure}
  \hfill
    \begin{subfigure}[t]{0.48\linewidth}
    \centering
    \includegraphics[width=\linewidth, trim=0 30 0 110, clip]{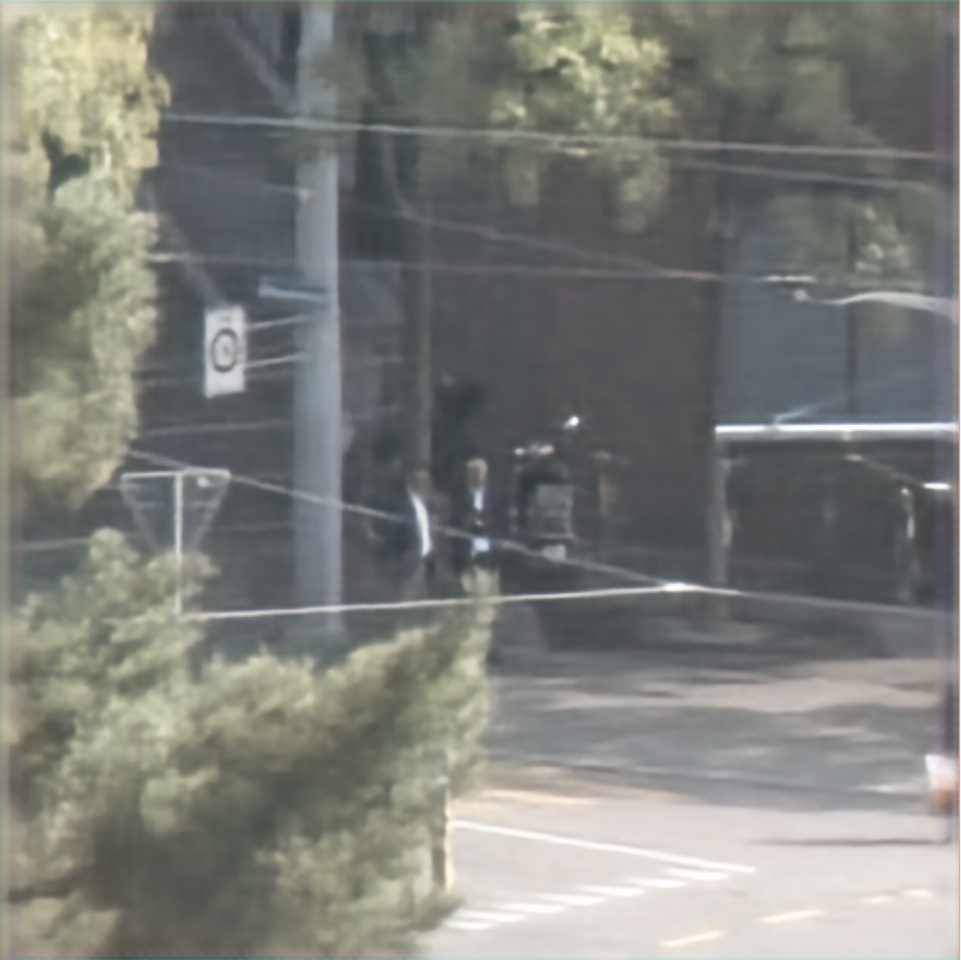}
    \caption{BSR: PSNR=52.55}
    \label{fig:ex_bsr}
  \end{subfigure}
  \par}

  \vspace{0.3em}

  {\centering
  \begin{subfigure}[t]{0.48\linewidth}
    \centering
    \includegraphics[width=\linewidth, trim=0 30 0 110, clip]{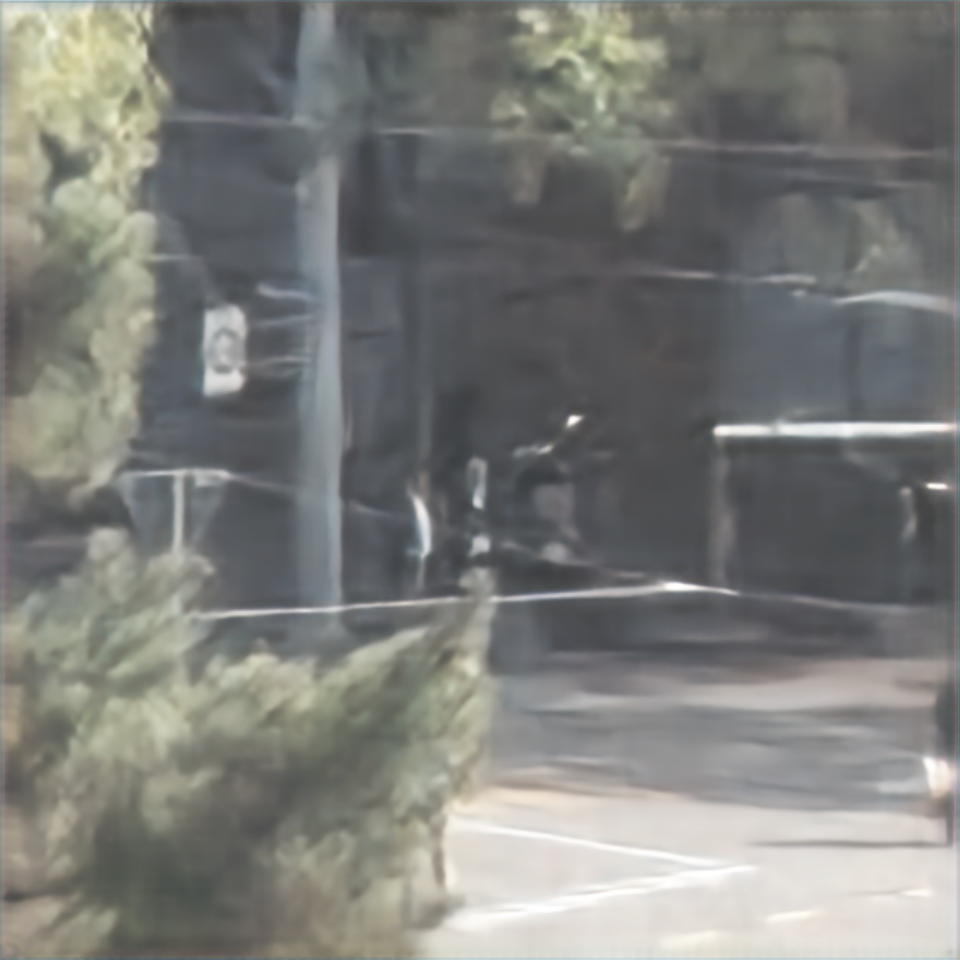}
    \caption{\textbf{PBSR}: PSNR=51.74}
    \label{fig:ex_pbsr}
  \end{subfigure}
  \hfill
  \begin{subfigure}[t]{0.48\linewidth}
    \centering
    \includegraphics[width=\linewidth, trim=0 30 0 110, clip]{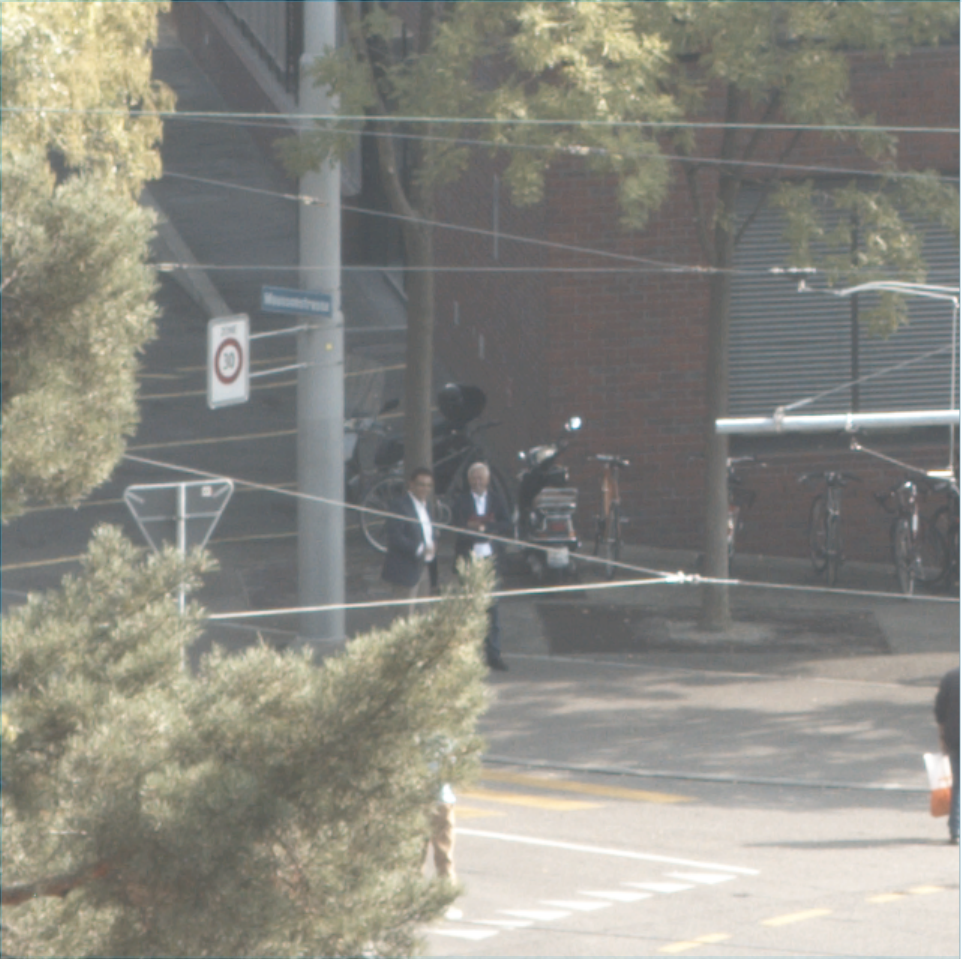}
    \caption{GT}
    \label{fig:ex_gt}
  \end{subfigure}
  \par}
    \caption{
    Effect of our PBSR. (a) SISR using an LR image.
    (b) BSR using 14 real burst LR frames. (c) PBSR from an LR image and 13 pseudo-burst LR frames. (d) Ground truth (GT).
    All results are reconstructed with the same DBSR~\cite{bhat2021deep} backbone, differing only in the input frames (single / real burst / pseudo-burst).
    }
  \vspace{-0.5em}
  \label{fig:teaser}
\end{figure}

Super-resolution (SR)~\cite{wang2020deep,liu2022video,liu2023blind} reconstructs high-resolution (HR) images from low-resolution (LR) images, with applications spanning smartphone photography~\cite{Chen_2019},
surveillance~\cite{zhang2010super}, and preprocessing for recognition~\cite{haris2021task,wang2020dual,kondo2024joint}.
Recent advances in deep learning
have improved SR quality~\cite{dong2015image,liang2021swinir}. Single-image SR (SISR)~\cite{Yang2014,timofte2017ntire,blau20182018} remains actively researched, yet it is inherently ill-posed because many plausible HR images correspond to a single LR image.
SISR methods therefore rely heavily on learned image priors, and the recovered textures and structures are often hallucinated. Fig.~\ref{fig:teaser}~(a) shows a typical failure, where the reconstruction is degraded by noise-like artifacts across the entire image.

To mitigate this ill-posedness, multi-frame SR (MFSR)~\cite{Tsai1984MultiframeIR} exploits complementary sub-pixel sampling across multiple LR frames of the same scene to reconstruct an SR image.
Within this framework, burst SR (BSR)~\cite{bhat2021deep} focuses on high-speed burst frames
captured by handheld cameras,
as shown in Fig.~\ref{fig:teaser}~(b).
However, MFSR, including BSR, requires multiple frames of the same scene. This requirement precludes use when (1) only a single image is available or (2) scene motion or large camera displacement prevents accurate frame alignment.

\begin{figure*}[t]
  \centering
  \begin{tikzpicture}
    \node[anchor=south west, inner sep=0] (img) at (0,0)
      {\includegraphics[width=\textwidth]{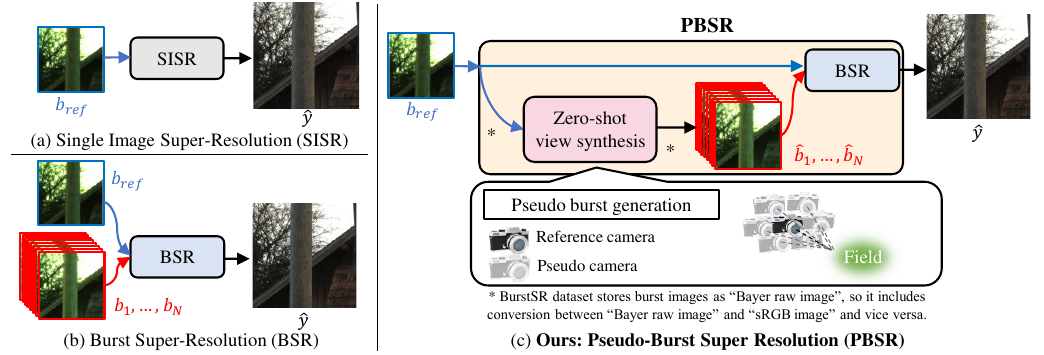}};
    \begin{scope}[x={(img.south east)}, y={(img.north west)}]

    \node[modlabel, anchor=west] at (0.760,0.92) {Sec.~\ref{subsec:overview}};

      \node[modlabel] at (0.63,0.76) {Sec.~\ref{subsec:pbsr_pb},~\ref{subsec:pbsr_viewpoints}};

      \node[modlabel] at (0.81,0.70) {Sec.~\ref{subsec:pbsr_integration},~\ref{subsec:nvs_gf},~\ref{subsec:pbsr_artifacts}};

    \end{scope}
  \end{tikzpicture}

  \vspace{0.3em}
    \caption{
    Conceptual overview of SISR/BSR baselines and our PBSR framework. PBSR generates pseudo-burst frames via zero-shot NVS, enabling BSR-style reconstruction from $b_{ref}$ under a single-image input condition.
    }
    \label{fig:overview}
\end{figure*}

These two families of methods leave a gap in the literature. SISR accepts a single image but cannot draw on multiple observations, while BSR exploits multiple observations but cannot accept a single image. Self-ensemble and related inference-time augmentation methods~\cite{timofte2016seven,lim2017enhanced,kimura2021understanding,deng2023efficient} attempt to bridge this gap. They augment a single input with geometric transformations such as shifts, rotations, flips, and scalings, and average the resulting SR outputs. However, such image-plane transforms provide no new viewpoint. The aim of this study is to bring BSR-style multi-frame reconstruction to the single-image input condition. To this end, we propose \emph{Pseudo-Burst Super-Resolution (PBSR)}, which, like SISR (Fig.~\ref{fig:overview}~(a)), needs only a single LR image (denoted by $b_{ref}$).

Instead of the real burst frames used in general BSR (Fig.~\ref{fig:overview}~(b)), PBSR synthesizes pseudo-burst frames (``Pseudo-burst frames'' in Fig.~\ref{fig:overview}~(c)) from $b_{ref}$.
To go beyond image-plane augmentations, we focus on motion parallax between burst frames.
Motion parallax is the depth-dependent shift in apparent object positions caused by a change in viewpoint: nearby objects shift more than distant ones.
Unlike image-plane transforms that merely reshuffle existing pixels, viewpoint shifts produce geometry-consistent sub-pixel appearance variations that can aid SR reconstruction.
In PBSR, pseudo-burst frames exhibiting such parallax are synthesized from $b_{ref}$ using zero-shot novel view synthesis (NVS)~\cite{zeronvs}, which reconstructs a 3D scene representation from a single image and renders it from nearby viewpoints.

Because the pseudo-burst frames are synthesized for explicitly sampled camera parameters rather than captured by a physical camera, PBSR can directly reuse these parameters for frame alignment. We exploit this property by introducing a geometry-aware alignment module. Specifically, by combining the camera parameters sampled during the NVS stage with an estimated depth map, we compute analytical flow maps. This explicit geometric guidance serves as a structural prior for frame alignment.

Beyond frame alignment, the generative nature of NVS introduces a separate challenge. Specifically, the synthesized pseudo-burst frames inevitably contain texture artifacts and geometric inconsistencies. To prevent the BSR backbone from erroneously aggregating these corrupted regions during the reconstruction phase, we introduce an artifact-aware adaptation. The backbone is fine-tuned using paired high-resolution images and their simulated pseudo-bursts. This training guides the network's attention mechanism to downweight artifact-dominated regions and selectively aggregate only reliable, complementary spatial cues. Note that the zero-shot property refers to the NVS stage. The BSR backbone itself is fine-tuned through this artifact-aware adaptation.

Our evaluation on the BurstSR dataset~\cite{bhat2021deep} focuses on two questions: whether pseudo-burst frames synthesized by zero-shot NVS improve reconstruction over the single-frame baseline under the same backbone, and how each component of PBSR contributes to this improvement.

Our contributions are threefold:
\begin{itemize}
\item A novel framework, PBSR, that synthesizes pseudo-burst frames from a single LR image via zero-shot NVS and enables BSR-style multi-frame reconstruction from a single image. PBSR recovers finer details than the single-frame baseline with the same backbone and can mitigate its hallucinations, as illustrated in Fig.~\ref{fig:teaser}~(c).
\item A pseudo-burst generation strategy that uses a 3D-aware neural scene representation adapted to $b_{ref}$, together with a viewpoint sampling scheme calibrated to real burst statistics. This strategy synthesizes motion parallax, including plausible content in newly visible regions, in the pseudo-burst frames.
\item An alignment module, \emph{NVS-Parametrized Geometric Flow (NVS-GF)}, which injects flow fields derived from the camera parameters specified during view synthesis. On average, NVS-GF performs comparably to the purely learned alignment. Its benefit appears on textured inputs, where it improves reconstruction robustness.
\end{itemize}

\section{Related Work}
\label{sec:related-work}

\subsection{Single-Image SR}
\label{subsec:sisr}

SISR has evolved from frequency-domain methods~\cite{Ji2009RobustWS} and exemplar-based approaches~\cite{Glasner2009,perez2016psyco} to deep learning-dominated techniques, including
post-upsampling architectures~\cite{CVPR2016RealShi}, progressive upsampling strategies~\cite{CVPR2017DeepLai}, iterative back-projection~\cite{Haris_2018},
and Transformer-based models~\cite{liang2021swinir}.
More recently, real-world SISR methods have exploited generative diffusion priors, such as StableSR~\cite{wang2024exploiting} and SeeSR~\cite{wu2024seesr}. These methods improve perceptual quality, yet they depend even more strongly on generative priors, and controlling hallucinated content remains an open issue.

Despite continuous progress, SISR reconstructs an HR image from a single LR observation, which makes the problem ill-posed and often forces these SISR approaches to rely on learned image priors.
This limitation is qualitatively illustrated by the single-image result in Fig.~\ref{fig:teaser}~(a).

\subsection{Multi-Frame SR and Burst SR}
\label{subsec:mfsr}
Classical MFSR methods include frequency-domain alignment~\cite{Tsai1984MultiframeIR}, image-prior-based reconstruction~\cite{hardie1998high}, and iterative back-projection~\cite{peleg1987improving}.
With burst frames,
Deep burst super-resolution (DBSR)~\cite{bhat2021deep} incorporates feature-space alignment with attention-based fusion mechanisms.
BSRT~\cite{luo2022bsrt} extracts global and context-aware features, which are then fused via optical flow-guided deformable alignment for robust reconstruction.
BIPNet~\cite{dudhane2022burst} focuses on early-stage complementary feature fusion to preserve fine details. Burstormer~\cite{dudhane2023burstormer} combines multi-scale features with back-projection~\cite{Haris_2018} and squeeze-and-excitation mechanisms~\cite{hu2018squeeze} to aggregate information across frames effectively.
More recent studies continue to advance BSR in several directions, including self-supervised training on real bursts~\cite{bhat2023self}, state-space architectures for sub-pixel fusion~\cite{di2025qmambabsr}, and diffusion-based reconstruction for improving perceptual quality~\cite{tokoro2024burst,kawai2025efficient}.

While these advances focus on algorithmic improvements, a few prior studies~\cite{UBSI2018,LBSNR2025} explore theory-driven selection of an informative subset from an already captured burst.
UBSI-based selection~\cite{UBSI2018} prunes frames by minimizing spectral interference to choose an ideally translated subset.
LBSNR-guided selection~\cite{LBSNR2025} uses multi-channel sampling and reference selection to maximize an SNR lower bound.
However, like general MFSR, these methods assume that multiple real frames have already been captured.
In contrast, our method addresses the scenario in which no burst is available and only a single image is provided. Instead of selecting an informative subset from a captured burst, we generate pseudo-burst frames from a single image to enable BSR-style multi-frame reconstruction without any real burst input.

\noindent
\noindent\textbf{Geometry-aware MFSR with Camera Parameters:}
Recent MFSR studies explicitly use camera parameters for precise alignment and fusion across multiple views. For instance, EpiMISR~\cite{epimisr2024} uses extrinsic and intrinsic camera parameters to compute epipolar correspondences between input images, thereby facilitating the fusion of the input images.
RefVSR~\cite{refvsr2022} exploits calibration between multi-camera streams to transfer high-frequency details from other input views to a reference view.
However, like the aforementioned frame selection methods~\cite{UBSI2018,LBSNR2025}, these methods require multiple captured frames whose camera parameters must be recovered through calibration. In contrast, our method synthesizes pseudo-burst frames via NVS, where intrinsic parameters are fixed by design and extrinsic parameters are explicitly specified at rendering time, so all camera parameters are known without calibration. The use of these known parameters as an alignment prior is discussed in Section~\ref{subsec:rw_pose_prior}.

\subsection{Novel View Synthesis}
\label{subsec:nvs}

Neural Radiance Fields (NeRF)~\cite{mildenhall2020nerf} achieves NVS by learning a 3D scene representation from multiple images of the target scene.
However, NeRF and its early extensions require dense multi-view input images of the target scene.
Recent few-shot and zero-shot NVS methods relax this requirement on the number of input images.
For example, DietNeRF~\cite{Jain_2021_ICCV} reduces the number of required input images through semantic-consistency constraints.
Zero-1-to-3~\cite{liu2023zero} and ZeroNVS~\cite{zeronvs} achieve zero-shot synthesis from a single image $b_{ref}$ by using diffusion priors to inpaint the unobserved parts of the 3D scene representation.
Subsequent studies further improve the multi-view consistency of single-image NVS~\cite{liu2024syncdreamer,voleti2024sv3d}.

These zero-shot NVS methods can render novel views from a single image, which allows our method to synthesize pseudo-burst frames from $b_{ref}$.
However, they inherently introduce generative artifacts, creating a distributional gap between the synthesized pseudo-burst frames and the real burst images expected by standard BSR models.
Prior work has shown that such a gap between generated images and downstream models degrades task performance unless explicitly addressed~\cite{haris2021task,kim2024sr4ir}.
Our NVS-based burst synthesis pipeline faces the same challenge, which requires a dedicated adaptation mechanism to bridge this domain gap (Section~\ref{subsec:pbsr_artifacts}).

\subsection{Pose as Conditioning, Optimization, and Alignment Prior}
\label{subsec:rw_pose_prior}

Camera pose has been exploited in three distinct roles that are closely related to this work.
First, pose-conditioned view synthesis uses camera extrinsics as an explicit condition to control and stabilize novel-view generation.
For example, Pose-Guided Diffusion enforces cross-view consistency by guiding the diffusion process with pose information, improving realism and consistency across viewpoints~\cite{tseng2023poseguided}.

Second, pose can be treated as an unknown to be estimated or refined.
NeRF-based localization and pose refinement optimize camera poses to better explain multi-view observations.
TrackNeRF performs bundle adjustment for NeRF under sparse and noisy views by using feature tracks~\cite{mai2024tracknerf}.
Nerfect Match refines poses by coupling NeRF rendering with 2D-3D correspondences, providing a robust matching-driven refinement scheme~\cite{zhou2024nerfectmatch}.
Related directions also combine NeRF rendering cues with feature matching for pose estimation~\cite{chen2024marryingnerf}.

Third, pose can serve as a geometric prior to derive correspondence fields for frame alignment, rather than to condition image generation.
GeoNet introduces a rigid-flow formulation that converts depth and camera motion into a geometry-induced flow, and learns a residual component to explain non-rigid effects and modeling errors~\cite{yin2018geonet}.

This paper shares the third viewpoint but targets a different use case.
The goal is not pose-conditioned synthesis, nor pose estimation or refinement.
Instead, the camera pose used in NVS is reused as a prior for burst alignment.
The pose prior provides a geometry-based initialization for warping, while the residual term in the burst SR backbone handles depth bias, occlusions, and synthesis inconsistencies.

\section{Problem Formulation}\label{sec:formulation}

\noindent\textbf{Super-Resolution:} An LR image $b_{ref}$ is obtained by applying a degradation operation $D$ to an HR image $y$, i.e., $b_{ref} = D(y)$, where $D$ is a composite function including degradation operations such as downsampling, blur, and noise addition.

An SISR model $S_{S}$ reconstructs $y$ from $b_{ref}$. The reconstructed SR image $\hat{y}$ is defined as follows:
\begin{equation}\label{eq:sisr}
\hat{y} = S_{S}(b_{ref}; \theta_{S}),
\end{equation}
where $\theta_{S}$ denotes the parameters of $S_{S}$.

On the other hand, a BSR model $S_{B}$ with the parameters $\theta_{B}$ reconstructs the SR image $\hat{y}$ from $b_{ref}$ plus $N-1$ burst frames $b_{1}, \dots, b_{N-1}$.
$\hat{y}$ is defined as follows:
\begin{equation}\label{eq:bsr}
\hat{y} = S_{B}(b_{ref}, b_{1}, \dots, b_{N-1}; \theta_{B}).
\end{equation}

\noindent\textbf{Novel View Synthesis:}
NVS synthesizes images at arbitrary new viewpoints from a set of observed input images. When a set of $M$ input images $\{ x_{obs}^{(j)} \}_{j=1}^{M}$ is given along with their respective extrinsic camera parameters $\{ P_{obs}^{(j)} \}_{j=1}^{M}$, an NVS model $G$ synthesizes an image $\hat{x}_{new}$ observed from a new viewpoint $P_{new}$
as $\hat{x}_{new} = G\left( \{ x_{obs}^{(j)}, P_{obs}^{(j)} \}_{j=1}^{M}, P_{new}; \theta_{N} \right)$, where $\theta_{N}$ denotes the parameters of $G$.
In particular, for zero-shot NVS ($M=1$), this reduces to:
\begin{equation}\label{eq:zero-shot-nvs}
\hat{x}_{new} = G\left( x_{obs}, P_{obs}, P_{new}; \theta_{N} \right).
\end{equation}

\section{Pseudo-Burst Super-Resolution}\label{sec:pbsr}

\subsection{Overview}\label{subsec:overview}

Our proposed PBSR $S_{P}$ is defined by the following function:
\begin{equation}\label{eq:pbsr-1}
\hat{y} = S_{P}(b_{ref}; \theta_{P}),
\end{equation}
where $\theta_{P}$ denotes the parameters of $S_{P}$. $S_{P}$ takes only $b_{ref}$ as input, as in the SISR setting defined in Eq.~\eqref{eq:sisr}.
The following subsections present the pipeline in processing order: pseudo-burst generation (Section~\ref{subsec:pbsr_pb}), viewpoint sampling (Section~\ref{subsec:pbsr_viewpoints}), BSR using synthesized pseudo-burst frames (Section~\ref{subsec:pbsr_integration}), and its geometric flow-prior module, NVS-GF (Section~\ref{subsec:nvs_gf}),
followed by artifact-aware adaptation (Section~\ref{subsec:pbsr_artifacts}).

\subsection{Pseudo-Burst Generation}
\label{subsec:pbsr_pb}

Pseudo-burst frames $\{\hat{b}_i\}_{i=1}^{N-1}$ are synthesized via zero-shot NVS from $b_{ref}$ following the formulation in Eq.~\eqref{eq:zero-shot-nvs}:
\begin{equation}\label{eq:pg-gen}
\hat{b}_i = G\!\big(b_{ref}, P_{ref}, P_{new}^{(i)}; \theta_{N}\big), \quad i=1,\dots,N-1,
\end{equation}
where $P_{ref}$ and $P_{new}^{(i)}$ are the camera parameters of the reference and the $i$-th novel viewpoint, respectively.
The effectiveness of the pseudo-burst frames depends on how the novel viewpoints $\{P_{new}^{(i)}\}$ are sampled. These viewpoints should introduce complementary sub-pixel variations while remaining close to the reference view to preserve NVS fidelity.
They should also ensure that even distant objects exhibit measurable sub-pixel shifts, which is important for outdoor scenes containing both near and far structures.
The sampling policy for $\{P_{new}^{(i)}\}$ is described next.

\subsection{Viewpoint Sampling}
\label{subsec:pbsr_viewpoints}

Novel viewpoints $\{P_{new}^{(i)}\}$ are sampled to synthesize pseudo-bursts that provide pixel shifts useful for SR while preserving NVS fidelity.
The sampling range must satisfy two requirements.
It should be large enough to produce non-negligible pixel shifts in far regions.
At the same time, it should remain small enough to avoid misalignment and geometric distortion.

\begin{figure}[t]
    \centering
    \includegraphics[width=\linewidth]{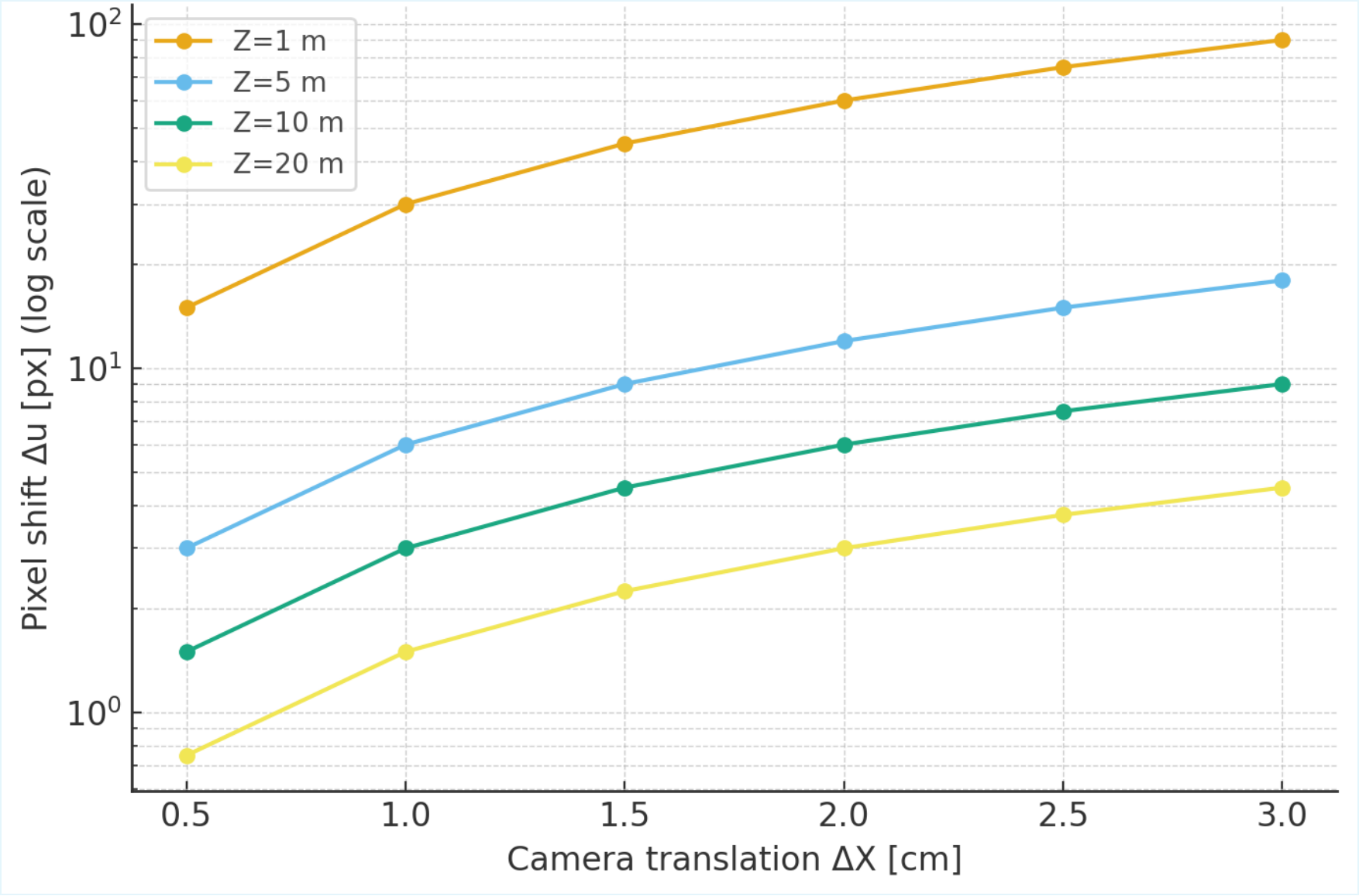}
    \vspace{-0.3em}
    \caption{Geometry-based sanity check of pixel shift versus camera translation under a pinhole approximation.
    The relation $\Delta u \approx f \Delta X / Z$ is plotted while sweeping depth $Z$ and translation $\Delta X$.
    A fixed focal length $f=3000$\,px is used.
    The plot shows that pixel shifts can drop by orders of magnitude as $Z$ increases, which motivates a non-trivial translation scale for far-field regions.}
    \label{fig:shift_translation_sweep}
    \vspace{-1.0em}
\end{figure}

Under a pinhole approximation, the pixel shift induced by a lateral camera translation $\Delta X$ roughly follows
\begin{equation}
\Delta u \approx f \frac{\Delta X}{Z},
\end{equation}
where $f$ is the focal length in pixel units and $Z$ is the scene depth.
This relation indicates that pixel shifts decrease rapidly as depth increases.
If $\Delta X$ is too small, far regions yield pixel shifts close to zero.
In that case, those regions contribute little additional sampling diversity to the pseudo-burst.
Fig.~\ref{fig:shift_translation_sweep} illustrates this dependence by sweeping $Z$ and $\Delta X$.

The relation above links the translation scale $\Delta X$ to the resulting pixel shift $\Delta u$.
However, $\Delta u$ also depends on depth through $\Delta u \propto 1/Z$.
A single $\Delta X$ cannot realize a fixed target pixel shift over a wide depth range.
In addition, with a single reference image, the metric depth scale is unknown.
Therefore, $\Delta X$ cannot be determined in metric units from a pixel-shift target without introducing a scale anchor.

\begin{figure}[t]
    \centering
    \includegraphics[width=\linewidth]{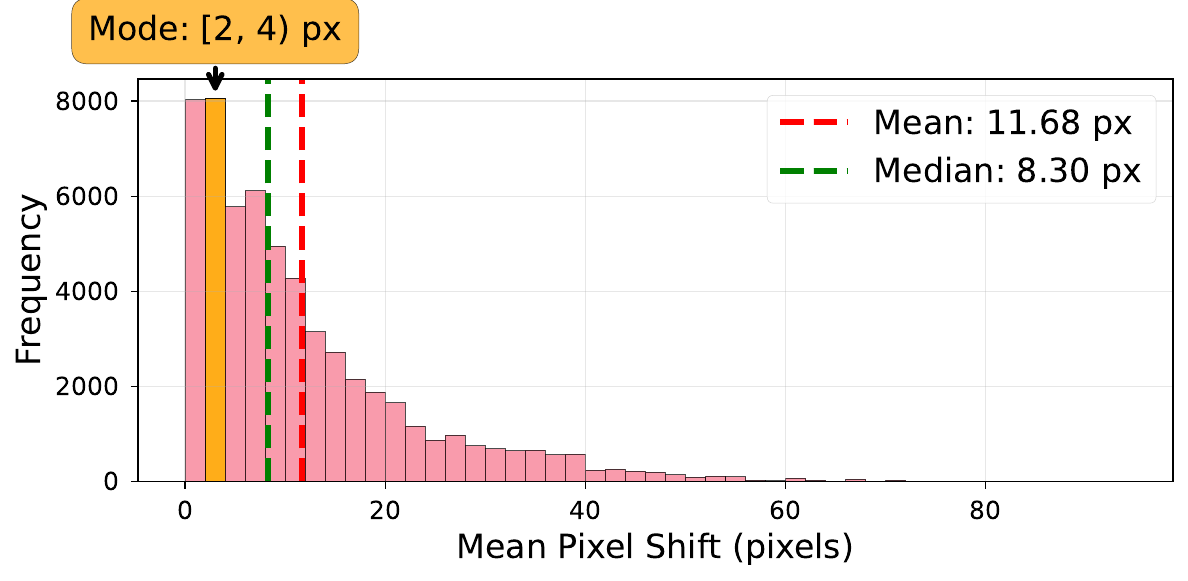}
    \vspace{-0.3em}
    \caption{Distribution of mean pixel shifts in the BurstSR dataset~\cite{bhat2021deep}.
    The histogram peaks in the range $[2,4)$\,px, indicating that most burst pairs involve only a few pixels of motion.}
    \label{fig:burstsr_shift_hist}
    \vspace{-1.0em}
\end{figure}

A representative depth $Z_{ref}$ is introduced as such an anchor.
For outdoor scenes, we consider a primary depth range of $10$--$20$\,m and place the anchor at the far end, $Z_{ref}=20$\,m, where pixel shifts are smallest.
A target pixel shift is then specified at the anchor depth as $\Delta u_{ref}\approx 4$\,px, which corresponds to $\Delta X \approx 30$\,mm for $f=3000$\,px, ensuring that even distant objects produce measurable shifts while keeping the viewpoints within a small neighborhood of $P_{ref}$.
A radial perturbation $\delta r \in [-0.005, 0.005]$ is additionally included to introduce mild forward-backward camera displacement along the optical axis (Algorithm~\ref{alg:zeronvs_sampling}). Its contribution to pixel shift is negligible ($\delta r / r_0 \approx 0.5\%$).
Finally, the target pixel shift is validated against real burst statistics.
Fig.~\ref{fig:burstsr_shift_hist} shows the distribution of mean pixel shifts in the BurstSR dataset~\cite{bhat2021deep} (mode: $[2,4)$\,px, median: $8.30$\,px).
At the anchor depth $Z_{ref}{=}20$\,m, the target shift is $\approx 4$\,px, while closer objects (e.g.\ $Z{=}10$\,m) produce shifts of $\approx 9$\,px, spanning the mode-to-median range of the dataset.
This confirms that the chosen sampling scale yields pixel shifts within the range where BSR models are known to operate on the BurstSR dataset.

\begin{algorithm}[t]
\setlength{\intextsep}{0pt}
\setlength{\textfloatsep}{0pt}
\caption{ZeroNVS Viewpoint Sampling}
\label{alg:zeronvs_sampling}
\DontPrintSemicolon
\SetKwInOut{Input}{Input}\SetKwInOut{Output}{Output}
\Input{Reference extrinsic $P_{ref}=(r_0,\theta_0,\phi_0)$;\; burst length $N-1$}
\Output{Sampled extrinsics $\{P_{new}^{(i)}\}_{i=1}^{N-1}$}
\For{$i=1$ \KwTo $N-1$}{
  $\delta\phi,\delta\theta \sim \mathcal{U}(-1.5^\circ,1.5^\circ)$,\;
  $\delta r \sim \mathcal{U}(-0.005, 0.005)$;\;
  $(r^{(i)},\theta^{(i)},\phi^{(i)}) \leftarrow
  (r_0,\theta_0,\phi_0) + (\delta r,\delta\theta,\delta\phi)$;\;
  $P_{new}^{(i)} \leftarrow \text{SphericalToExtrinsic}(r^{(i)},\theta^{(i)},\phi^{(i)})$
}
\end{algorithm}

\begin{figure*}[t]

    \centering

    \begin{subfigure}{0.7\linewidth}
      \centering
      \begin{tikzpicture}
        \node[inner sep=0] (base) {\includegraphics[width=\linewidth]{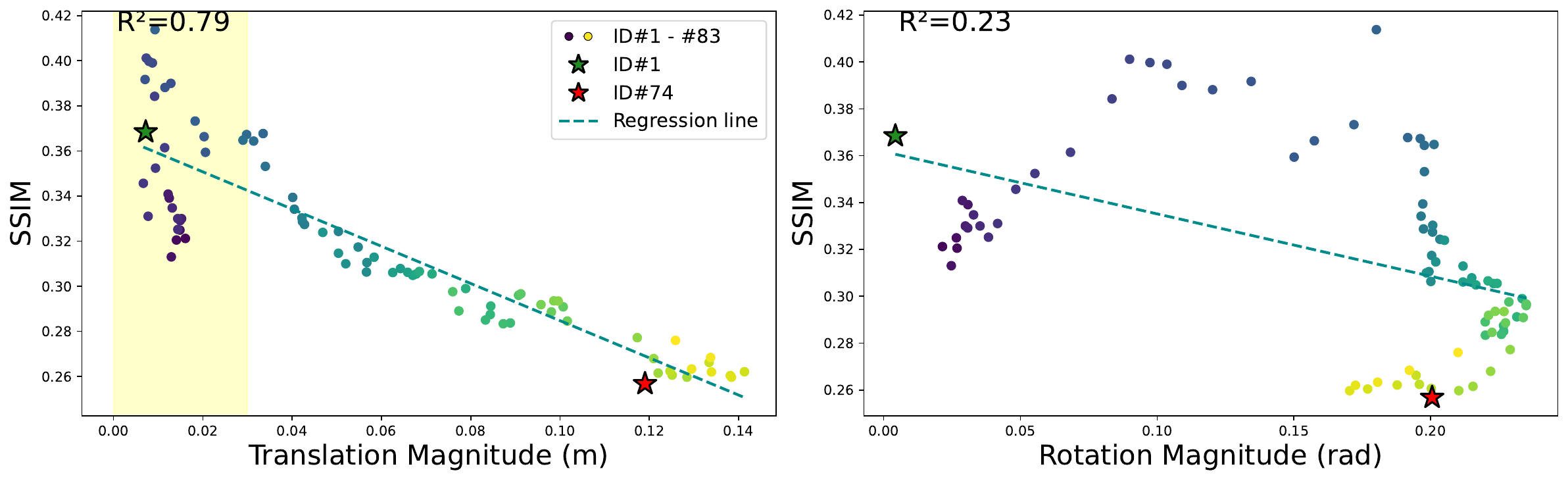}};
      \end{tikzpicture}

      \vspace{-0.2em}
      \caption{SSIM vs.\ translation (left) and rotation (right)}
      \label{fig:2dplots}
    \end{subfigure}\hfill%
    \begin{subfigure}{0.29\linewidth}
        \centering
        \includegraphics[width=\linewidth]{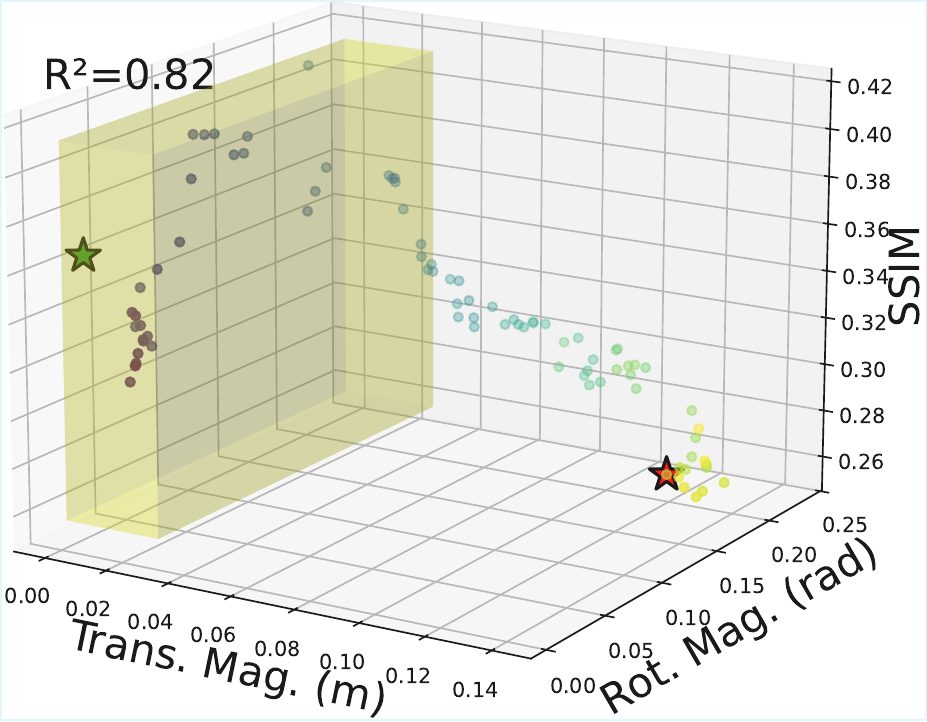}
        \caption{SSIM vs.\ translation/rotation}
        \label{fig:3dplot}
    \end{subfigure}\hfill%

    \vspace{0.1em}

    \hspace{0.04\textwidth}
    \begin{subfigure}{0.17\textwidth}
        \centering
        \includegraphics[width=\linewidth]{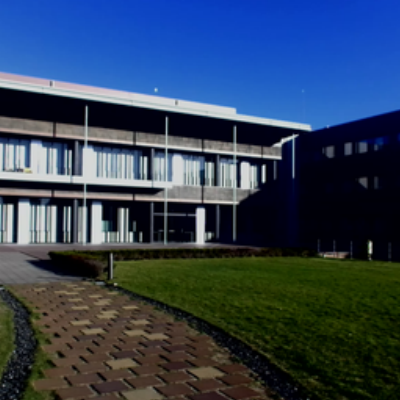}
        \caption{Reference image}
        \label{fig:ref}
    \end{subfigure}
    \hspace{0.04\textwidth}
    \begin{subfigure}{0.34\textwidth}
        \centering
        \begin{tikzpicture}
            \node[inner sep=0] {\includegraphics[width=\linewidth]{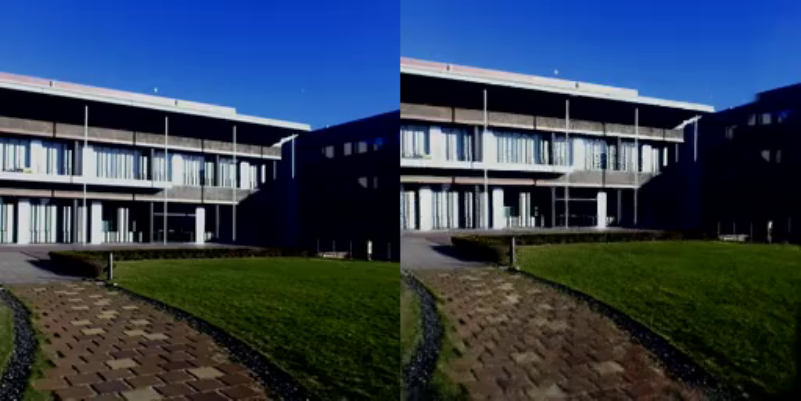}};
            \node[anchor=north east, fill=white, text=black, font=\footnotesize] at (3.08,1.52) {SSIM=0.368};
        \end{tikzpicture}
        \caption{\greenstar ID\#1 ET (left) and prediction (right)}
        \label{fig:best}
    \end{subfigure}
    \hspace{0.04\textwidth}
    \begin{subfigure}{0.34\textwidth}
        \centering
        \begin{tikzpicture}
            \node[inner sep=0] {\includegraphics[width=\linewidth]{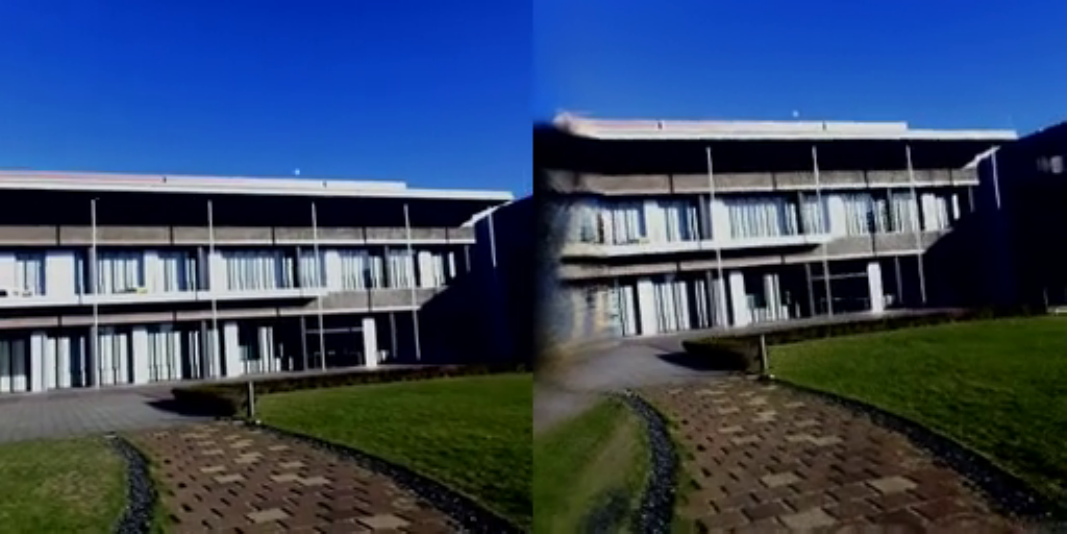}};
            \node[anchor=north east, fill=white, text=black, font=\footnotesize] at (3.08,1.52) {SSIM=0.257};
        \end{tikzpicture}
        \caption{\redstar ID\#74 ET (left) and prediction (right)}
        \label{fig:worst}
    \end{subfigure}\\
    \vspace{0em}
    \caption{
    ZeroNVS~\cite{zeronvs} analysis on our ZED Mini capture.
    (a)--(b) show SSIM as a function of camera translation and rotation magnitudes, with points color-coded by frame index.
    The dashed line denotes a least-squares linear fit; $R^2$ is the coefficient of determination.
    The yellow-highlighted region indicates the range of extrinsic camera-parameter variations used for burst capture.
    (c) Reference image, (d) initial frame (ID\#1), and (e) failure case (ID\#74).
    }
        \label{fig:nvs-psnr}
    \vspace{-1em}
\end{figure*}

To examine how variations in extrinsic camera parameters affect reconstruction fidelity, we conduct a ZeroNVS~\cite{zeronvs} analysis on
a handheld outdoor sequence captured with a ZED Mini camera.
We use the first frame (Fig.~\ref{fig:nvs-psnr}~(c)) as the reference image, and use the subsequent 83 frames as evaluation targets (ET) for the ZeroNVS renderings.
Each point in Fig.~\ref{fig:nvs-psnr}~(a)--(b) corresponds to one evaluation frame and is color-coded according to its frame ID, forming a smooth gradient that represents the sequential camera extrinsics.
For each plot, we report the coefficient of determination $R^2$ of a least-squares linear regression (dashed line).

In Fig.~\ref{fig:nvs-psnr}~(a) (left), a clear negative correlation is observed between SSIM and translation magnitude ($R^2{=}0.79$), indicating that larger positional changes lead to gradual degradation in reconstruction fidelity.
In contrast, Fig.~\ref{fig:nvs-psnr}~(a) (right) shows that rotational changes have a weaker influence on SSIM ($R^2{=}0.23$).
Fig.~\ref{fig:nvs-psnr}~(b) plots SSIM as a function of both translation and rotation, confirming that translation is the dominant factor ($R^2{=}0.82$).
Meanwhile, large extrinsic-parameter variations, such as ID\#74, lead to pose-dependent misalignment and structural artifacts (Fig.~\ref{fig:nvs-psnr}~(e)), a known limitation of zero-shot NVS models at large viewpoint displacements~\cite{zeronvs,chen2024cascade,zheng2024free3d}.
The translation scale $\Delta X \approx 30$\,mm derived above falls within the yellow-highlighted region in Fig.~\ref{fig:nvs-psnr}~(a) and (b), where SSIM remains high, confirming that the adopted sampling range preserves stable pseudo-burst image quality.

\subsection{BSR Using Pseudo-Burst Frames}
\label{subsec:pbsr_integration}

Given $\{\hat{b}_i\}_{i=1}^{N-1}$ synthesized according to Eq.~\eqref{eq:pg-gen}, a BSR model $S_{B}$ in Eq.~\eqref{eq:bsr} becomes applicable under the single-image input condition by composing $G$ and $S_{B}$:
\begin{equation}
\hat{y} = S_{B}\!\big(b_{ref}, \hat{b}_1(b_{ref}), \dots, \hat{b}_{N-1}(b_{ref}); \theta_{B}\big).
\end{equation}

Since all inputs originate from $b_{ref}$, the formulation still satisfies the single-image input condition defined in Eq.~\eqref{eq:pbsr-1}.

\begin{figure}[t]
    \centering
    \includegraphics[width=1.00\linewidth]{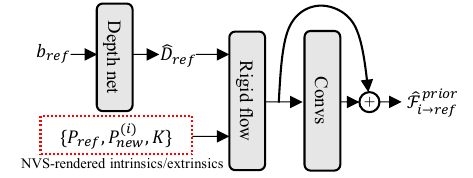}\\
    \vspace{0.3em}
    \caption{NVS-GF module. This module is integrated into the alignment stage of the BSR backbone.}
    \vspace{-0.5em}
    \label{fig:nvs_gf}
\end{figure}

\subsection{NVS-Parametrized Geometric Flow (NVS-GF)}
\label{subsec:nvs_gf}

Most BSR models $S_B$ estimate optical flow $\hat{\mathcal{F}}_{i\to ref}$ to align input frames.
In PBSR, pseudo-burst frames are rendered with known camera extrinsics from the NVS process.
These camera parameters provide a pose prior for alignment.
NVS-GF (Fig.~\ref{fig:nvs_gf}) converts this pose prior into a pixel-wise flow prior and provides it to the BSR backbone as an explicit initialization.

\noindent\textbf{Terminology:}
The term \emph{rigid flow} $\hat{\mathcal{F}}^{rig}_{i\to ref}$ refers to a pixel-wise flow field computed under a static-scene assumption.
It is obtained by projecting each reference-view pixel into a novel view using the depth map $\hat{D}_{ref}$ and the camera parameters (Eq.~\eqref{eq:rigid_flow}).
NVS-GF outputs a \emph{geometric flow prior} $\hat{\mathcal{F}}^{prior}_{i\to ref}$ by adding a learned residual that compensates for depth errors and NVS artifacts (Eq.~\eqref{eq:prior_flow}).
Given a depth map $\hat{D}_{ref}$ estimated from $b_{ref}$ via monocular depth estimation (Section~\ref{subsec:impl_details}) and camera parameters $(P_{ref}, P^{(i)}_{new})$, the rigid flow $\hat{\mathcal{F}}^{rig}_{i\to ref}$ is computed as
\begin{equation}
\hat{\mathcal{F}}^{rig}_{i\to ref}(\bm{p})
=
\pi\!\Big(
K\,
P^{(i)}_{{new}}\,
P_{ref}^{-1}\,
\hat{D}_{ref}(\bm{p})\,
K^{-1}\bm{p}
\Big)
- \bm{p},
\label{eq:rigid_flow}
\end{equation}
where $\bm{p}=(u,v,1)^{\top}$ is a homogeneous pixel coordinate in the reference view
and $K$ denotes the intrinsic camera matrix.
Specifically, $K^{-1}\bm{p}$ back-projects $\bm{p}$ into a normalized camera ray,
$\hat{D}_{ref}(\bm{p})\,K^{-1}\bm{p}$ lifts this ray to a 3D point at depth $\hat{D}_{ref}(\bm{p})$ in the reference camera frame,
$P_{ref}^{-1}$ transforms the point into the world frame,
$P^{(i)}_{new}$ re-projects it into the $i$-th novel-view camera frame,
and $\pi(\bm{x})=(x/z,\,y/z,\,1)^{\top}$ for $\bm{x}=(x,y,z)^{\top}$ applies perspective division to yield pixel coordinates.
Monocular depth estimation yields either relative depth, which lacks metric scale, or absolute depth, which still contains estimation errors.
In either case, the resulting rigid flow is approximate.
The rigid flow $\hat{\mathcal{F}}^{rig}_{i\to ref}$ in Eq.~\eqref{eq:rigid_flow} can be biased due to depth errors and artifacts in NVS-rendered pseudo-burst frames.
A lightweight correction network takes $\hat{\mathcal{F}}^{rig}_{i\to ref}$ as input and predicts a residual $\Delta\hat{\mathcal{F}}^{nvs}_{i\to ref}$.
The geometric flow prior output by NVS-GF is defined as
\begin{equation}
\hat{\mathcal{F}}^{prior}_{i\to ref}
=
\hat{\mathcal{F}}^{rig}_{i\to ref}
+
\Delta\hat{\mathcal{F}}^{nvs}_{i\to ref}.
\label{eq:prior_flow}
\end{equation}

\noindent\textbf{Backbone refinement:}
The BSR backbone further refines the prior flow through its internal alignment module.
Let $\Delta\hat{\mathcal{F}}^{bsr}_{i\to ref}$ denote the backbone-predicted refinement.
The final alignment flow used for warping is defined as
\begin{equation}
\hat{\mathcal{F}}_{i\to ref}
=
\hat{\mathcal{F}}^{prior}_{i\to ref}
+
\Delta\hat{\mathcal{F}}^{bsr}_{i\to ref}.
\label{eq:final_flow}
\end{equation}
In Eq.~\eqref{eq:final_flow}, $\Delta\hat{\mathcal{F}}^{nvs}_{i\to ref}$ mainly compensates for geometry-related bias caused by depth errors and NVS artifacts.
The term $\Delta\hat{\mathcal{F}}^{bsr}_{i\to ref}$ accounts for the remaining non-idealities and improves final alignment accuracy.

\begin{figure}[t]
    \centering
    \includegraphics[width=1.00\linewidth]{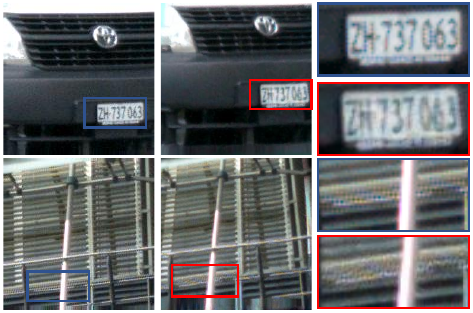}
        ~~~
        \footnotesize
        ~\hspace*{4mm} (a) Real \hspace*{13mm} (b) Pseudo \hspace*{6mm}
        (c) Enlarged view\\
        \vspace{1em}
    \caption{Examples of artifacts in pseudo-burst frames, including distortion in license plates (first row) and variations in noise and blur (second row).}
    \vspace{-1em}
    \label{fig:pb_artifact}
\end{figure}

\subsection{Artifact-Aware Adaptation}
\label{subsec:pbsr_artifacts}

Pseudo-burst frames may contain generative artifacts (Fig.~\ref{fig:pb_artifact}) that can be erroneously exploited during BSR reconstruction.
Fine-tuning the BSR backbone on the pseudo-burst frames mitigates this effect.
The training objective is inherited from the underlying BSR model $S_B$; no artifact-specific loss is introduced (details in Section~\ref{subsec:impl_details}).
Specifically, given LR-HR pairs from an SR training set, each LR image is regarded as $b_{ref}$ and used for synthesizing its pseudo-burst frames, as described in Section~\ref{subsec:pbsr_pb} and Section~\ref{subsec:pbsr_viewpoints}. Pseudo-burst frames, which include NVS- and image-dependent artifacts, are fed into the BSR model to reconstruct the SR image. The BSR model is then fine-tuned by minimizing the loss between its SR output and the HR image, which realizes the artifact-aware adaptation.
This adaptation guides attention to reliable regions and prevents artifacts from being incorrectly exploited during reconstruction.
Sections~\ref{subsec:comparison} and~\ref{subsec:fusion_weights} report the empirical validation of this adaptation and the scaling behavior as the number of pseudo-burst frames increases.

\section{Experimental Results}\label{sec:experiment}

We evaluate PBSR on the BurstSR validation set~\cite{bhat2021deep}.
Following the dataset protocol, each evaluation sample is an $80{\times}80$ RAW crop with 14 burst frames.
The upscaling factor is $4{\times}$.
PSNR, SSIM, and LPIPS follow the official aligned evaluation of BurstSR~\cite{bhat2021deep}. This protocol spatially and photometrically aligns each output to the ground truth before scoring. The alignment compensates for the camera differences between the burst frames and the ground truth.
In the PBSR setting, we assume only a single captured RAW frame is available as input, and treat it as the reference frame $b_{ref}$.
\subsection{Implementation Details}\label{subsec:impl_details}
ZeroNVS operates in sRGB, whereas DBSR takes BGGR RAW as input.
To bridge this mismatch, we convert $b_{ref}$ from RAW to sRGB using the processing in BIPNet~\cite{dudhane2022burst}, and convert each synthesized sRGB frame $\hat{b}_i$ back to RAW using the method of Nam \textit{et al.}~\cite{nam2022learning}.
For NVS-GF, we estimate a relative depth map $\hat{D}_{ref}$ from $b_{ref}$ using DepthFM~\cite{gui2025depthfm}.
Since the depth is defined up to an affine ambiguity, we normalize it by its median and apply learnable scale and shift parameters to obtain a consistent depth proxy for geometric flow computation.
The intrinsic matrix $K$ is taken from the Samsung Galaxy S8 metadata used in BurstSR.
These auxiliary modules are used only for synthesizing and aligning pseudo-burst observations. Throughout our comparisons, the SR backbone is kept fixed to DBSR to isolate the effect of additional (pseudo) observations.

\noindent\textbf{Artifact-aware adaptation (additional training):}
We report two PBSR variants depending on whether the SR backbone is adapted to pseudo-burst inputs.
(i) \textit{Zero-shot} PBSR (denoted as PBSR (z/s)) directly uses the publicly trained DBSR weights without any fine-tuning on pseudo-bursts.
(ii) \textit{Adapted} PBSR fine-tunes the DBSR backbone to reduce the domain gap between real bursts and pseudo-bursts, which is caused by NVS and sRGB-to-RAW conversion artifacts.
For this adaptation, we randomly select 646 training samples from the BurstSR training set and fine-tune for 80 epochs.
During fine-tuning, we keep the reference frame $b_{ref}$ as the original captured frame and replace the remaining frames with synthesized pseudo-burst frames $\{\hat{b}_i\}$.
This training protocol matches the PBSR inference condition (single captured reference) while exposing the backbone to pseudo-burst artifacts in the auxiliary observations.
The loss, optimizer, and learning-rate schedule follow the DBSR training protocol.

\noindent\textbf{Preprocessing for sRGB-to-RAW conversion:}
To further suppress image degradation introduced by the sRGB-to-RAW conversion, we fine-tune the conversion network of Nam \textit{et al.}~\cite{nam2022learning} for 120 epochs.
This fine-tuning is applied only to the conversion module, and its output is used for both PBSR (z/s) and the adapted PBSR variant.
Other hyperparameters, loss functions, and optimizers follow Nam \textit{et al.}~\cite{nam2022learning}.

\begin{table}[t]
    \centering
    \caption{Quantitative comparison of DBSR~\cite{bhat2021deep} and PBSR on the BurstSR dataset~\cite{bhat2021deep}. DBSR-$N$ denotes the same DBSR backbone with $N$ input frames (DBSR-1: single-frame inference baseline). \underline{Underlined values} indicate that PBSR performs better than DBSR-1 under the same DBSR backbone.}\label{tab:quant-result}
    \begin{tabular}{lccc}
    \toprule
    Methods            & PSNR $\uparrow$ & SSIM $\uparrow$ & LPIPS $\downarrow$ \\
    \midrule
    DBSR-1~\cite{bhat2021deep}             & 44.695          & 0.971           & 0.053              \\
    DBSR-2~\cite{bhat2021deep}             & 46.112          & 0.977           & 0.040              \\

    DBSR-14~\cite{bhat2021deep}            & 47.698          & 0.984           & 0.029              \\
    \midrule
    PBSR-2 (z/s)       & 44.065          & 0.970           & \underline{0.048}              \\
    PBSR-14 (z/s)      & 41.887          & 0.954           & 0.071              \\
    \midrule
    PBSR-2             & \underline{45.140} & \underline{0.974} & \underline{0.040} \\
    PBSR-14            & \underline{45.633} & \underline{0.976} & \underline{0.044} \\
    \bottomrule
    \end{tabular}
    \vspace{-0.5em}
\end{table}

\subsection{Comparison with Single-Image Input and Burst Models}\label{subsec:comparison}

To assess how closely PBSR approaches the reconstruction quality of BSR while retaining the single-image convenience of SISR, we compare PBSR with DBSR models trained on the BurstSR dataset. Since DBSR accepts a variable number of burst frames $N$ at inference time, we compare PBSR and DBSR by varying $N\in[2, 14]$. Additionally, we evaluate DBSR with $N{=}1$ as a single-frame baseline in the single-image input setting (i.e., the SISR setting). This $N{=}1$ baseline is not a dedicated SISR model; it is used to isolate the effect of the number of observations under a fixed DBSR backbone. All models are denoted as $\mathcal{M}-N$ for model name $\mathcal{M}$. Evaluation metrics include PSNR, SSIM~\cite{ssim}, and LPIPS~\cite{lpips}.

Quantitative evaluation results are shown in Table~\ref{tab:quant-result}. Even PBSR-2, which uses only two frames, achieves better results than DBSR-1 across all evaluation metrics. PSNR and SSIM improve slightly from PBSR-2 to PBSR-14, but LPIPS slightly deteriorates. Artifact-aware adaptation brings a substantial gain over PBSR (z/s).
The degradation of PBSR (z/s) as the number of pseudo-burst frames increases suggests a quality gap between real bursts and pseudo-bursts, mainly caused by NVS generation bias and the sRGB-to-RAW conversion. Without additional training, this gap is not compensated for, and extra pseudo-burst frames can introduce misleading cues.
Neither PBSR-2 nor PBSR-14 reaches the PSNR and SSIM of DBSR-2.

\begin{figure}[t]
    \centering
    \begin{subfigure}[b]{\linewidth}
        \centering
        \includegraphics[width=\linewidth]{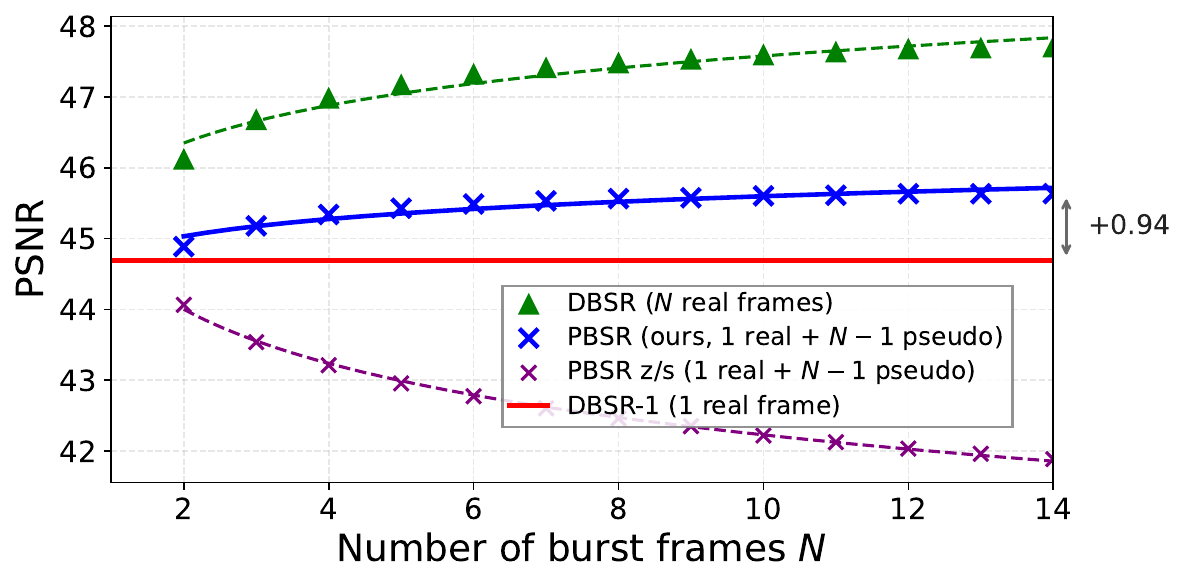}\\
        \vspace{-0.2em}
        \caption{PSNR}
        \label{fig:psnr}
    \end{subfigure}
    \vspace{0.2em}  %
    \begin{subfigure}[b]{\linewidth}
        \centering
        \includegraphics[width=\linewidth]{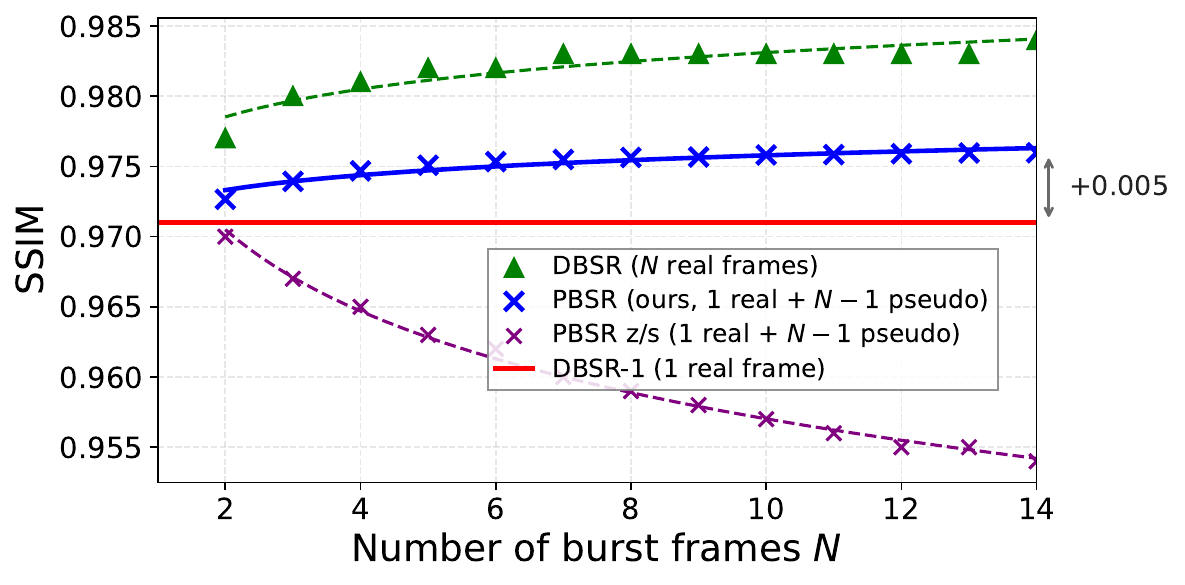}\\
        \vspace{-0.2em}
        \caption{SSIM}
        \label{fig:ssim}
    \end{subfigure}
    \vspace{0.2em}  %
    \begin{subfigure}[b]{\linewidth}
        \centering
        \includegraphics[width=\linewidth]{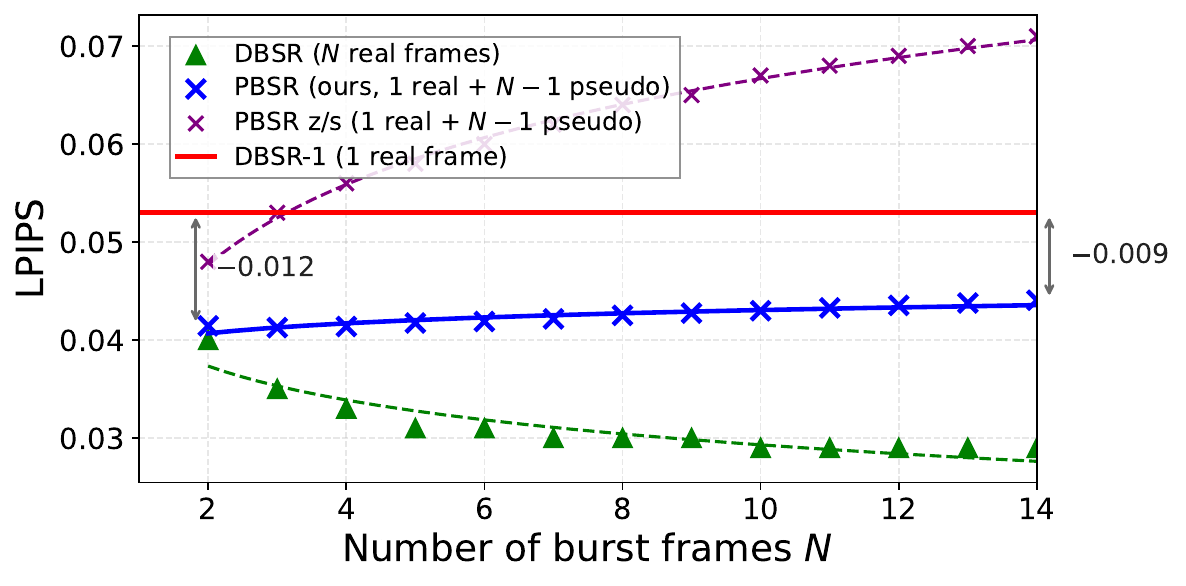}\\
        \vspace{-0.2em}
        \caption{LPIPS}
        \label{fig:lpips}
    \end{subfigure}\\
    \caption{PSNR, SSIM, and LPIPS as a function of burst length $N$.
DBSR uses $N$ real captured burst frames; PBSR (proposed) and PBSR (z/s) use only one real reference frame $b_{ref}$ and $N{-}1$ synthesized pseudo-burst frames.
Dashed lines show logarithmic-scale quadratic interpolation.}
    \vspace{-0.5em}
    \label{fig:metrics_comparison}
\end{figure}

Fig.~\ref{fig:metrics_comparison} reports metrics for burst lengths from 2 to 14 frames. With artifact-aware adaptation (PBSR), PSNR and SSIM consistently increase as burst length grows, while LPIPS gradually worsens. This suggests that extra pseudo-burst frames improve structure but inject texture biases from NVS and color biases from the sRGB-to-RAW conversion, which LPIPS penalizes. In contrast, the zero-shot variant PBSR (z/s) degrades as burst length increases, indicating that artifact-aware adaptation is crucial for exploiting NVS-generated bursts.

\newif\ifvertical
\verticaltrue   %

\begin{figure*}[t]
    \small
    \centering
    \includegraphics[width=\textwidth]{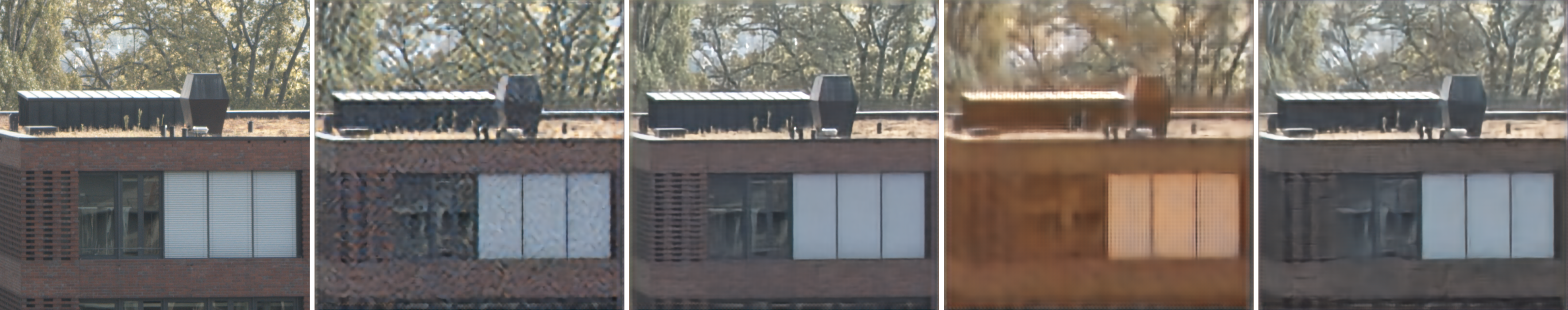}
    \includegraphics[width=\textwidth]{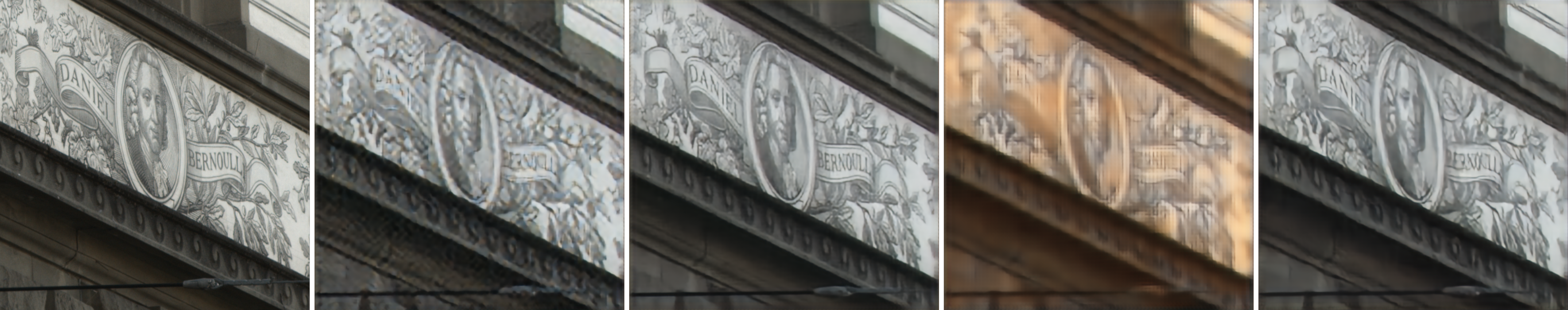}
    \includegraphics[width=\textwidth]{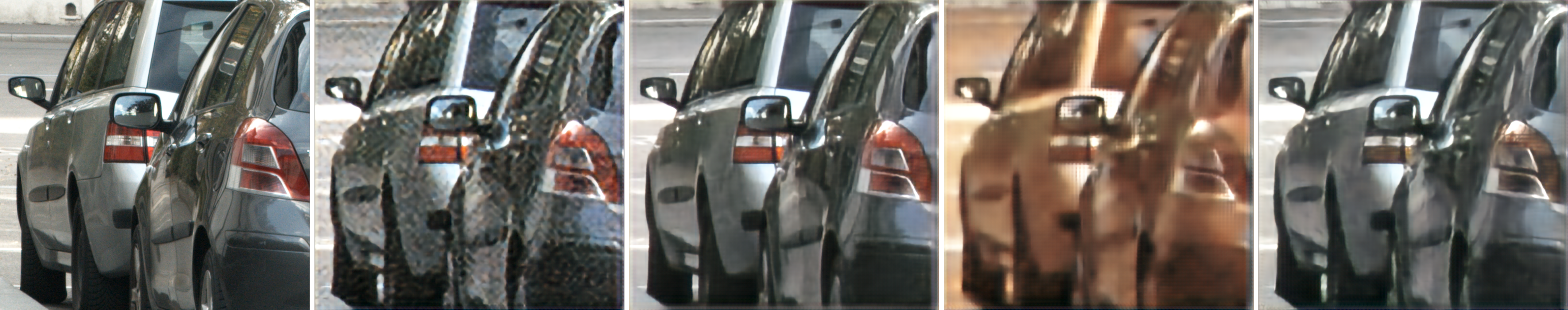}
    \includegraphics[width=\textwidth]{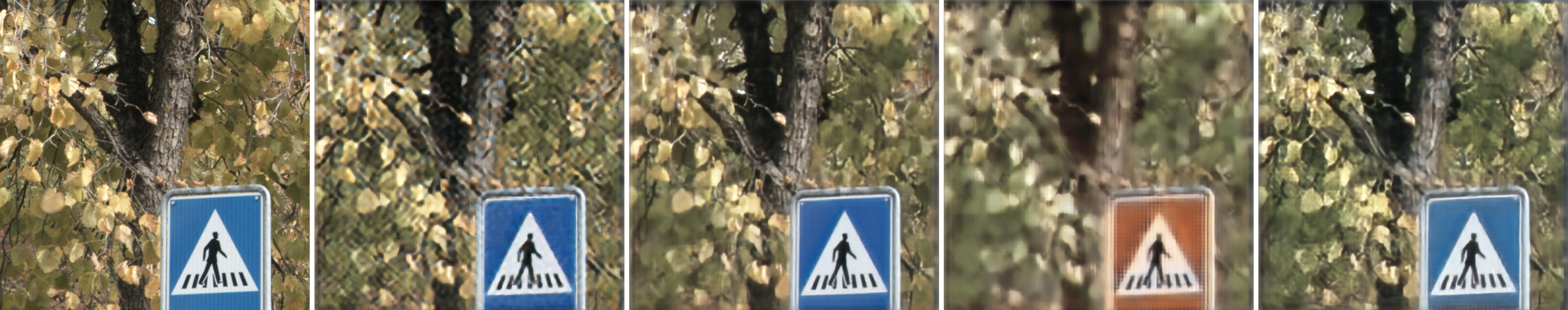}\\
        ~~~
        \footnotesize
        ~\hspace*{-9mm} (a) Ground truth \hspace*{9mm} (b) DBSR-1~\cite{bhat2021deep} \hspace*{10mm}
        (c) DBSR-14~\cite{bhat2021deep} \hspace*{8mm}
        (d) PBSR-14 (z/s) \hspace*{10mm}
        (e) PBSR-14\\
        \vspace{0.5em}
    \caption{Qualitative $\times 4$ SR comparisons on the BurstSR dataset~\cite{bhat2021deep}.
    }
    \label{fig:qual-result}
    \vspace{-1.3em}
\end{figure*}

Fig.~\ref{fig:qual-result} presents qualitative $\times 4$ SR results.
DBSR-1 (Fig.~\ref{fig:qual-result}~(b)) shows pattern-like artifacts and blurred details, while DBSR-14 (Fig.~\ref{fig:qual-result}~(c)) reconstructs sharper structures.
PBSR-14 (z/s) (Fig.~\ref{fig:qual-result}~(d)) suffers from a noticeable brownish color shift and strong blur across all examples.
This degradation is reduced in PBSR-14 with artifact-aware adaptation (Fig.~\ref{fig:qual-result}~(e)), where the color shift is largely suppressed and edge sharpness improves.

\begin{table}[t]
    \centering
    \caption{Evaluation of pseudo-burst generation in the adapted PBSR-14 on the BurstSR validation set. ``Shift+Rot.'' indicates geometric pseudo-image generation with parallel translation and rotation transformations. Best values are shown in \sota{bold red}.}\label{tab:pbs-result}
    \vspace{-0.5em}
    \setlength{\tabcolsep}{2.5pt}%
    \begin{tabular}{lccc}
    \toprule
    Methods            & PSNR $\uparrow$ & SSIM $\uparrow$ & LPIPS $\downarrow$ \\
    \midrule
    Shift+Rot.      & 44.985          & 0.972           & 0.075              \\
    \textbf{Pseudo-burst generation}           &  \sota{45.633} & \sota{0.976} & \sota{0.044} \\
    \bottomrule
    \end{tabular}
    \vspace{-1em}
\end{table}

This color shift originates from the pseudo-burst generation pipeline rather than from the BSR backbone. The synthesized views exhibit no systematic color bias in sRGB space. The sRGB-to-RAW conversion~\cite{nam2022learning}, however, introduces a consistent warm-direction bias into the pseudo-burst frames (R excess in 70\%, G deficiency in 77\%, and B deficiency in 88\% of the evaluation crops, measured per crop in the RAW domain). Because the zero-shot model assigns 69\% of its fusion weight to the pseudo-burst frames at $N{=}14$ (Section~\ref{subsec:fusion_weights}), this bias propagates to the output and accumulates as the number of pseudo-burst frames grows. Measured in the display domain, 98\% of the 882 evaluation crops exhibit a perceptible color cast (CIELAB magnitude above one~\cite{mahy1994evaluation,mokrzycki2011color}). Among these casts, 77\% lie in the warm (brownish) direction. Artifact-aware adaptation suppresses the shift by re-weighting the fusion toward the reference frame (from 31\% to 55\% of the total weight), reducing the cast to the level of the real-burst baselines. Indeed, matching the color statistics of the pseudo-burst frames to the reference frame removes the cast almost entirely and recovers 2.18~dB in the aligned PSNR for the zero-shot model. This simple matching alone, however, still falls short of the adapted model. Since the aligned evaluation already compensates global color differences, this recovery mainly reflects restored fusion quality rather than the color correction itself. These results indicate that the inter-frame color inconsistency is a major cause of both the color shift and the blur in Fig.~\ref{fig:qual-result}~(d), rather than the fusion architecture itself.
The contrast between (d) and (e) suggests that the zero-shot backbone is sensitive to pseudo-burst artifacts introduced by NVS and the sRGB-to-RAW conversion, and that adaptation helps the backbone cope with these artifacts during reconstruction.
Although PBSR-14 remains inferior to DBSR-14 due to the lack of real multi-frame observations, it produces sharper outputs than DBSR-1 while requiring only a single captured reference frame.

\subsection{Analysis of Pseudo-Burst Generation}

This section analyzes why NVS-based pseudo-burst generation improves reconstruction quality.
One hypothesis is that NVS produces geometry-consistent viewpoint variations, providing more useful multi-view cues than a single reference image $b_{ref}$.

An alternative hypothesis is that the improvement mainly comes from using multiple inputs itself, similar to inference-time data augmentation~\cite{kimura2021understanding, deng2023efficient}.
To test this alternative, a 2D geometric baseline is introduced for comparison.

Specifically, pseudo frames are generated by applying image-plane transformations to $b_{ref}$.
Translation is independently sampled in the range $[-4,4]$ pixels for each axis, and rotation is sampled in the range $[-1,1]$ degrees.
This baseline increases the number of inputs but does not introduce geometry-consistent parallax, since it is limited to 2D transformations.

Table~\ref{tab:pbs-result} shows that pseudo-burst generation using NVS consistently outperforms Shift+Rot. across all metrics.
This result indicates that NVS-based pseudo-bursts provide more useful multi-view cues than simple 2D perturbations, and this contributes to the reconstruction gains of PBSR.

\subsection{Analysis of NVS-GF}

\begin{table}[t]
    \centering
    \caption{Evaluation of NVS-GF in the adapted PBSR-14 on the BurstSR validation set. ``w/'' applies NVS-GF, while ``w/o'' removes it; all other modules are kept fixed.}\label{tab:nvs-rf-result}
    \vspace{-0.5em}
    \begin{tabular}{lccc}
    \toprule
    Methods            & PSNR $\uparrow$ & SSIM $\uparrow$ & LPIPS $\downarrow$ \\
    \midrule
    w/o NVS-GF      & 45.557          & 0.976           & 0.043             \\
    w/ NVS-GF           & 45.633 & 0.976 & 0.044 \\
    \bottomrule
    \end{tabular}
\end{table}

\begin{figure}[t]
    \centering
            \includegraphics[
            width=0.99\linewidth,
            trim=0.5cm 0 0 0, clip   %
        ]{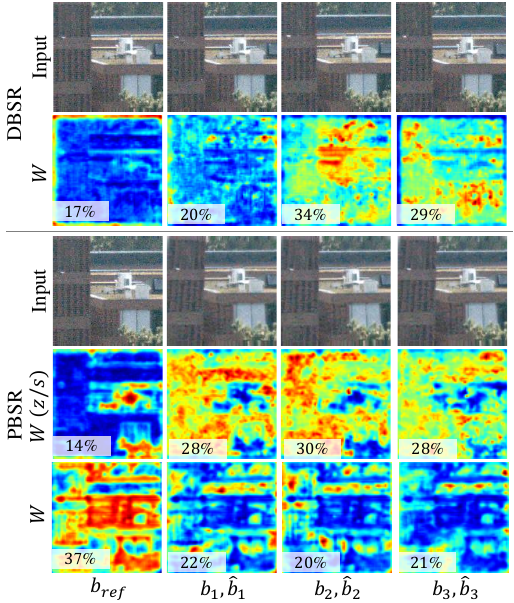}\\
    \vspace{0em}
    \caption{Fusion-weight $W$ visualization.
The figure is divided into two groups: DBSR (upper two rows) and PBSR (lower three rows).
Each group shows the input burst frames (\textit{Input} row) and the corresponding fusion-weight heatmaps.
For PBSR, two heatmap rows are shown: $W(z/s)$ (without artifact-aware adaptation) and $W$ (with adaptation).
In PBSR, the three non-reference inputs ($\hat{b}_1, \hat{b}_2, \hat{b}_3$) are pseudo-burst frames.}
    \label{fig:fusion_weight}
\end{figure}

The effect of incorporating NVS-GF into the alignment module is evaluated in Table~\ref{tab:nvs-rf-result}. NVS-GF yields performance comparable to the purely learned alignment across all three metrics. A per-image paired analysis over the 882 evaluation crops shows a mean difference of +0.041~dB with a 95\% confidence interval of $[+0.003, +0.078]$. This analysis re-scores both conditions in a single new evaluation session. Its mean therefore differs from the difference of the single-run scores in Table~\ref{tab:nvs-rf-result} (0.076~dB). Both values are of the same order as the run-to-run variation of the evaluation. The difference is, however, not uniform across scenes. When the crops are grouped by the gradient energy of the reference frame, NVS-GF improves the majority of the textured crops. It also lifts the lower tail of the per-image PSNR distribution (+0.33~dB at the 1st percentile). The cost is a marginal PSNR decrease on flat, low-texture crops. We therefore regard NVS-GF as a robustness-oriented initialization rather than a source of gains in the average metrics.

\subsection{Analysis of Fusion Weights in DBSR and PBSR}\label{subsec:fusion_weights}

We analyze how DBSR, PBSR (z/s), and adapted PBSR fuse information when the input burst length is fixed to four frames ($N{=}4$).
In this analysis, DBSR uses four real burst frames, whereas PBSR (z/s) and adapted PBSR use one real reference frame $b_{ref}$ and three synthesized pseudo-burst frames.
Fusion weights control how feature maps extracted from individual burst frames are combined. They are obtained from the attention-based fusion module in the DBSR backbone and are summarized as a single-channel heatmap for visualization and per-frame contribution analysis.
Following the official DBSR implementation, the 512-channel fusion-weight tensor is reduced to a single-channel heatmap by channel-wise aggregation, and each frame's contribution is computed from the summed weights.

As shown in Fig.~\ref{fig:fusion_weight}, DBSR and PBSR (z/s) assign large fusion weights to non-reference frames.
In DBSR, this behavior is natural because all inputs are real burst frames.
In PBSR (z/s), however, the auxiliary frames are pseudo-bursts and can include NVS and sRGB-to-RAW conversion artifacts.
The strong reliance on these auxiliary frames (86\% in Fig.~\ref{fig:fusion_weight}) is consistent with the discoloration and blur observed in Fig.~\ref{fig:qual-result}.

In contrast, adapted PBSR increases the fusion weight on the real reference frame $b_{ref}$ (37\% in Fig.~\ref{fig:fusion_weight}), while keeping non-trivial weights on pseudo-burst frames (about 20\% each).
This suggests that training encourages the backbone to treat $b_{ref}$ as more reliable, without imposing an explicit constraint for reference-only fusion.
The same tendency holds at $N{=}14$, where the pseudo-burst frames receive 69\% of the total weight in PBSR (z/s) and adaptation raises the reference-frame weight from 31\% to 55\%.

\section{Conclusion}\label{sec:conclusion}

In this work,
we propose PBSR, which brings BSR-style multi-frame reconstruction to the single-image input setting by synthesizing pseudo-burst frames with zero-shot NVS. PBSR combines NVS-based viewpoint sampling, geometry-aware alignment via NVS-GF, and artifact-aware adaptation of a BSR backbone. Experiments on BurstSR show that PBSR improves in PSNR, SSIM, and LPIPS over the $N{=}1$ single-frame baseline with the same DBSR backbone.
They also show that NVS-based pseudo-burst generation outperforms simple geometric perturbations. NVS-GF performs comparably to the purely learned alignment on average and improves robustness on textured inputs.

\noindent\textbf{Limitations and future work:}
PBSR has limitations. First, all SR experiments use the BurstSR dataset, the DBSR backbone, and ZeroNVS. NVS-GF is evaluated in the adapted PBSR-14 setting. Validating the framework with other backbones, NVS models, and datasets is a natural next step. Second, with artifact-aware adaptation, adding pseudo-burst frames raises PSNR and SSIM but slightly worsens LPIPS (Fig.~\ref{fig:metrics_comparison}). A similar divergence between pixel-wise and perceptual metrics is well known in SR~\cite{blau2018perception}. In PBSR, the extra frames add structural cues, while their texture and color biases are penalized by LPIPS. Both PBSR-2 and PBSR-14 still improve over DBSR-1 in all three metrics (Table~\ref{tab:quant-result}). The burst length can then be chosen according to the target metric. Third, the quality of pseudo-burst frames remains the main bottleneck. NVS introduces texture artifacts and geometric inconsistencies (Fig.~\ref{fig:pb_artifact}), and the sRGB-to-RAW conversion adds a warm-direction color bias (Section~\ref{subsec:comparison}). Artifact-aware adaptation reduces these effects, yet PBSR-14 still trails DBSR-2, which uses two real frames. Stronger NVS priors and improved viewpoint sampling may narrow this gap and are left for future work.

\section*{Declarations}

\begin{itemize}
\item \textbf{Funding:} This research was supported by JSPS KAKENHI Grant Number 23K24874.
\item \textbf{Competing interests:} The authors declare that they have no competing interests.
\item \textbf{Ethics approval and consent to participate:} Not applicable.
\item \textbf{Consent for publication:} Not applicable.
\item \textbf{Code availability:} The code is not publicly available at this time.
\item \textbf{Data availability:} The data are not publicly available at this time.
\item \textbf{Materials availability:} Not applicable.
\item \textbf{Author contributions:} Y. Kondo conceived the idea and conducted the experiments. N. Ukita supervised the project and provided critical feedback. All authors reviewed and approved the final manuscript.
\end{itemize}

\section*{Acknowledgments}\label{sec:acknowledgments}
This research was supported by JSPS KAKENHI Grant Number 23K24874. We express our gratitude to Kyotaro Tokoro of Toyota Technological Institute for his assistance with BSR implementation and prior research investigation.

\small
\bibliography{main}

\section*{Author Biographies}

\begin{biography}{\includegraphics[width=64pt]{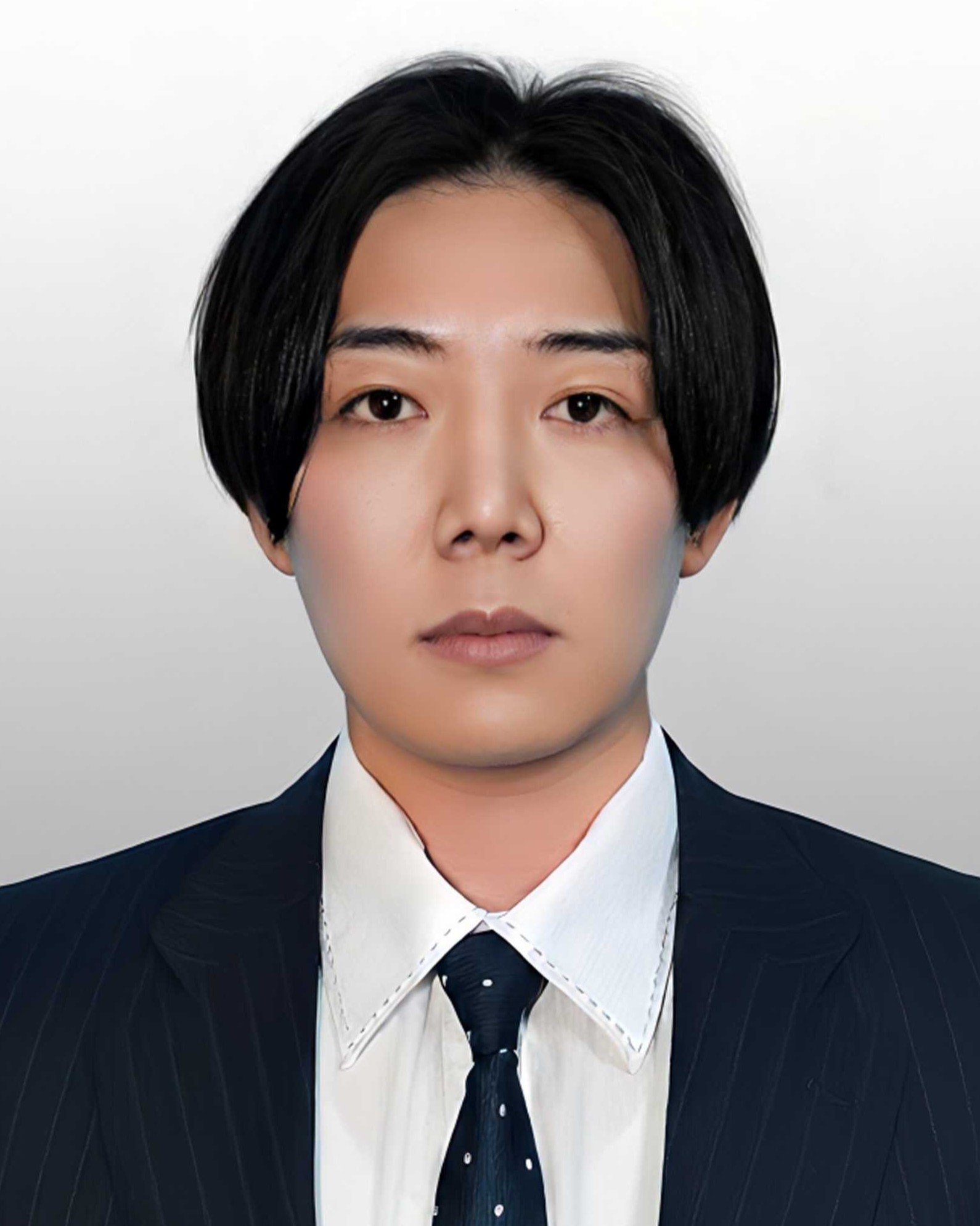}}{\author{Yuki Kondo} received the B.E. degree from Toyota Technological Institute, Japan, in 2022. Since 2022, he has been with the Frontier Research Center, Toyota Motor Corporation, Japan. Since 2025, he has also been a Ph.D. student at Toyota Technological Institute under the supervision of Norimichi Ukita. His research interests include super-resolution, low-level vision, small object recognition, and robotic vision. He received the Best Practical Paper Award at MVA2021 and the Encouragement Award from IPSJ SIG-CVIM in 2024. He served as a Technical Event Chair of MVA2023 and MVA2025, and is currently serving as a Technical Event Chair of MVA2027 and a Tutorial Chair of ACCV2026. He has served as a reviewer for CVPR, AAAI, IEEE TMM, CVIU, Advanced Robotics, and IEICE TIS.}
\end{biography}

\begin{biography}{\includegraphics[width=64pt]{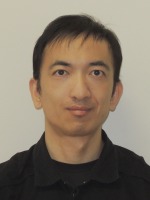}}{\author{Norimichi Ukita} received the B.E. and M.E. degrees in information engineering from Okayama University in 1996 and 1998, respectively, and the Ph.D. degree in Informatics from Kyoto University in 2001. From 2001 to 2016, he was an assistant professor and an associate professor at Nara Institute of Science and Technology. In 2016, he became a professor at Toyota Technological Institute. He was a research scientist with Precursory Research for Embryonic Science and Technology, Japan Science and Technology Agency, from 2002 to 2006, and a visiting research scientist at Carnegie Mellon University from 2007 to 2009. Currently, he is also an adjunct professor at Toyota Technological Institute at Chicago. His research interests include object detection, tracking, and human pose estimation.}
\end{biography}

\end{document}